\documentclass[aps,prx,twocolumn,nofootinbib,superscriptaddress]{revtex4}

\usepackage{amsmath,amssymb,bm}
\usepackage{graphicx}
\usepackage{fancybox,color}
\usepackage[justification=centerlast]{caption, subfig} 
\usepackage{float}

\usepackage[utf8]{inputenc}
\usepackage{bbold}

\usepackage[normalem]{ulem}

\usepackage{braket} 

\newlength{\myboxL}

\def\ra{\rightarrow}

\def\e#1{{\color{blue} \uline{#1}}}

\def\hc{\text{h.c.}}

\def\>{\rangle}
\def\<{\langle}

\renewcommand{\Im}{\mbox{Im}}
\def\v#1{{\bm{#1}}}
\def\op#1{\mathbf{#1}}
\def\d{\partial}

\def\h{\hat}
\def\t{\widetilde}

\def\s{{}^\dagger}

\def\p{\partial}

\def\h{\hat}
\def\t{\widetilde}
\def\s{{}^\dagger}
\def\sc{{}^*}
\def\1{\mathbf{1}}

\def\la{\lambda}
\def\d{\delta}
\def\eps{\varepsilon}

\def\k{\v k}

\def\K{\v K}
\def\dK{\d \K}
\def\tK{\t{\K}}
\def\suma{\sum_l}
\def\da{\v d_l}

\def\e#1{\v e_{#1}}
\def\te#1{\t{\v e}_{#1}}

\def\teps{\underline{\t{\bm \eps}}}
\def\leps{\underline{{\bm \eps}}}
\def\lA{\mathsf{A}}
\def\lB{\mathsf{B}}
\def\lEdge{\mathsf{C}}
\def\lCon{\mathsf{P}}

\def\lb{\lambda}
\newcommand{\un}[1]{\ensuremath{\,\mathrm{#1}}}

\begin{document}

\title{Current flow paths in deformed graphene:\\ from quantum transport to classical trajectories in curved space}

\author{Thomas Stegmann}
\email[e-mail:]{stegmann@icf.unam.mx } 
\affiliation{Fakult{\"a}t f{\"u}r Physik, Universit{\"a}t Duisburg-Essen, Duisburg, Germany}
\affiliation{Instituto de Ciencias Físicas, Universidad Nacional Autónoma de México, Cuernavaca, México}

\author{Nikodem Szpak}
\email[e-mail:]{nikodem.szpak@uni-due.de}
\affiliation{Fakult{\"a}t f{\"u}r Physik, Universit{\"a}t Duisburg-Essen, Duisburg, Germany}

\date{\today}

\begin{abstract}
In this work we compare two fundamentally different approaches to the electronic transport in deformed graphene:
a) the condensed matter approach in which current flow paths are obtained by applying the non--equilibrium Green's function (NEGF) method to the tight--binding model with local strain,
b) the general relativistic approach in which classical trajectories of relativistic point particles moving in a curved surface with a pseudo--magnetic field are calculated.
The connection between the two is established in the long--wave limit via an effective Dirac Hamiltonian in curved space.
Geometrical optics approximation, applied to focused current beams, allows us to directly compare the wave and the particle pictures.
We obtain very good numerical agreement between the quantum and the classical approaches for a fairly wide set of parameters, improving with the increasing size of the system.
The presented method offers an enormous reduction of complexity from irregular tight--binding Hamiltonians defined on large lattices to geometric language for curved continuous surfaces.
It facilitates a comfortable and efficient tool for predicting electronic transport properties in 
graphene nanostructures with complicated geometries.
Combination of the curvature and the pseudo--magnetic field paves the way to new interesting transport phenomena such as bending or focusing (lensing) of currents depending on the shape of the deformation. 
It can be applied in designing ultrasensitive sensors or in nanoelectronics.
\end{abstract}

\maketitle

\section{Introduction}

It is a well known fact in relativistic particle physics that electromagnetic and gravitational fields couple differently to particles and antiparticles --- the first distinguishes their opposite charges, the second treats equivalently their identical masses.
In graphene, both these phenomena can be simulated by the action of magnetic and pseudo--magnetic fields as well as the influence of curvature on the dynamics of particle--like excitations and holes.
Continuous models of elastically deformed graphene, describing the excitations effectively by a two--dimensional Dirac equation for massless fermions coupled to an artificial magnetic field, have been extensively studied in the literature in recent years, see e.g.
\cite{Katsnelson+Geim-GrapheneRipples, CastroNeto+Guinea+Novoselov+Geim-GrapheneReview, Guinea+Katsnelson+Vozmediano-GrapheneBumpMidgapState, Guinea-Ripples, Juan+Cortijo+Vozmediano-AharonovBohm, Guinea+Katsnelson+Geim-LandauLevelsInStrainedG, Guinea+Geim+Katsnelson+Novoselov-BendingG, Guinea+CastroNeto-StrainInduced300T, Pereira+CastroNeto-StrainEngineering, Guinea-StrainEngineering, MuchaKruczynski+Flako-PseudomagneticFDistrib, Pereira-StrainedGraphene, Covaci-NanopillarsStrainedGraphene, Moldovan+Peeters-GrapheneStrainRevisited, Naumis-HomogStrainedGraphene, UniaxialStrainInShapedGrapheneStripe} for a representative selection or \cite{Cortijo+Juan+Vozmediano+Guinea-StrainedGraphene} for the most recent overview of the topic. 
Further geometric aspects, related to the Dirac equation in curved space, such as coupling to the effective metric leading to position and direction dependent Fermi velocity, 
have not yet attracted as much attention
\cite{VozKatsGui-review, Vozmediano-Ripples, Vozmediano-GaugeFields+Curvature, Graphene-curvature-hopping, Juan+Vozmediano-SpaceDepFermiVelG, Juan+Manes+Vozmadiano-GaugeFieldsFromStrain, TransportStrainedGraphene, Naumis-NonuniformlyStrainedGraphene}.
Especially, the electronic transport in the presence of curvature centers \cite{TransportStrainedGraphene} still lacks detailed qualitative and quantitative results.
The key question elaborated here, on the relation between the electric currents and classical trajectories, has been addressed before in the presence of pure pseudo--magnetic \cite{CirculatingCurrents-DeformedGraphene, Covaci-WaveScatteringStrainedGraphene}, real magnetic \cite{Stegmann+Lorke-MagnetotransportGraphene}, combined pseudo--magnetic and real magnetic \cite{Real+PseudomagneticF-Graphene} or pure electric fields \cite{Katsnelson-KleinTunneling+Caustics+Lens-Graphene} without taking into account further influence from the curvature. 
However, as we will show below, the metric effect can be as relevant as that of the pseudo--magnetic field.

In this work, we study the effect of curvature on the current flow lines in elastically deformed graphene sheets.
Using a tight--binding model, we first apply the non--equilibrium Green's function (NEGF) method to obtain the current flow paths in the presence of a localized deformation. 
Then, utilizing the specific dispersion relation of graphene at low energies, we turn to the continuous approximation for long wavelengths, which leads us to the two--dimensional Dirac equation coupled to an effective pseudo--magnetic field and an attractive curvature.
In order to visualize the action of both factors,
we consider coherent current beams and apply the eikonal approximation.
This simplifies the approach to the semiclassical particle picture in which the current lines turn out to be
geodesics for relativistic charged massless particles moving in a curved two--dimensional surface in the presence of a magnetic field.
We compare the numerically obtained current flows from the NEGF calculations with the classical geodesic lines.
We find very good agreement, improving with the increasing size of the system, for a wide range of parameters of experimental interest.
The combination of both emergent forces -- curvature and the pseudo--magnetic field -- paves the way to interesting transport phenomena such as bending or focusing (lensing) of currents depending on the shape of the deformation. 
It can find technological applications in designing ultrasensitive sensors or in nanoelectronics.
The presented analogy facilitates a comfortable and efficient tool for calculating electronic transport properties in newly designed graphene nanostructures with complicated geometries.
It offers an enormous reduction of complexity from hopping Hamiltonians defined on large deformed lattices to a semiclassical geometric language for the description of curvature effects in continuous surfaces.
We conclude with the proposition of a geometrical lens for currents flowing through the deformed graphene.
Similar electron optics phenomena in graphene have been proposed in \cite{Falko-VeselagoLens, VeselagoLens+Current, ElectronOpticsG, Katsnelson-KleinTunneling+Caustics+Lens-Graphene} but the focusing was caused by $p$-$n$ junctions and by the electric potential.

Here, we do not take into account either the electron--electron interaction \cite{Pereira-ElectronElectronInteractionG}, the recombination of electronic density due to deformation  \cite{Fasolino+Katsnelson-GrapheneRipples} or the relaxation of the lattice structure to its minimal elastic energy \cite{Katsnelson-GrapheneBumpMidgapState+Relaxation}, leaving these topics open to further investigation.

\section{The effective Dirac equation in curved space}

\subsection{Tight--binding model for deformed graphene} \label{sec:Dirac-from-lattice}

We begin with the tight--binding Hamiltonian describing the hopping of electronic excitations in the lowest band
(in units in which $\hbar = e = 1$)
\begin{equation} \label{Ham-hopping-op}
   {\op{\h H}}
   = \sum_{\v n\in \lA} \sum_{l=1}^3 \left( t_{\v n,l}\ \h b\s_{\v n + \da}\ \h a_{\v n} + t^*_{\v n,l}\ \h a\s_{\v n}\ \h b_{\v n + \da} \right)
\end{equation}
The creation and annihilation operators $\h a\s_{\v n}, \h a_{\v n}$ and $\h b\s_{\v n+\da}, \h b_{\v n+\da}$ belong to the sublattices $\lA$ and $\lB$, respectively.
Vector $\v n$ runs over all points of the sublattice $\lA$ and the three vectors $\da$ point along the links to the nearest neighbors in the sublattice $\lB$. 
We choose
$\v d_1=(\frac1{2},\frac{\sqrt{3}}{2})\, d_0$, $\v d_2=(-1,0)\, d_0$, $\v d_3=(\frac1{2},-\frac{\sqrt{3}}{2})\, d_0$,
where $d_0$ is the interatomic distance in pristine graphene.
The hopping parameters in pristine graphene 
are all equal, $t_{\v n,l} = t_0$, and become position ($\v n$) and direction ($l$) dependent in deformed graphene which reflects  modified tunneling probabilities between neighboring atoms.
In the following, energies are measured in multiples of $t_0 = 2.8 \un{eV}$ and distances in multiples of $d_0 = 0.142 \un{nm}$, if not given explicitly.

Since we do not take into account any interaction effects between the electrons, the second quantized Hamiltonian \eqref{Ham-hopping-op} effectively reduces to a one-particle sector and can be written in a basis of localized states%
\footnote{In the course of this paper, there is no need to further distinguish between the $\lA$ and $\lB$ sublattices.}
$|\psi_{\v n}\>$ for $\v n\in \lA\cup \lB$
as
\begin{equation} \label{eq:Ham-hopping}
  \op{H}  = \sum_{\v n\in \lA} \sum_{l=1}^3  t_{\v n,l}\ |\psi_{\v n + \da}^\lB\>\<\psi_{\v n}^\lA| + \hc
\end{equation}

\begin{figure}
  \includegraphics[width=0.44\linewidth]{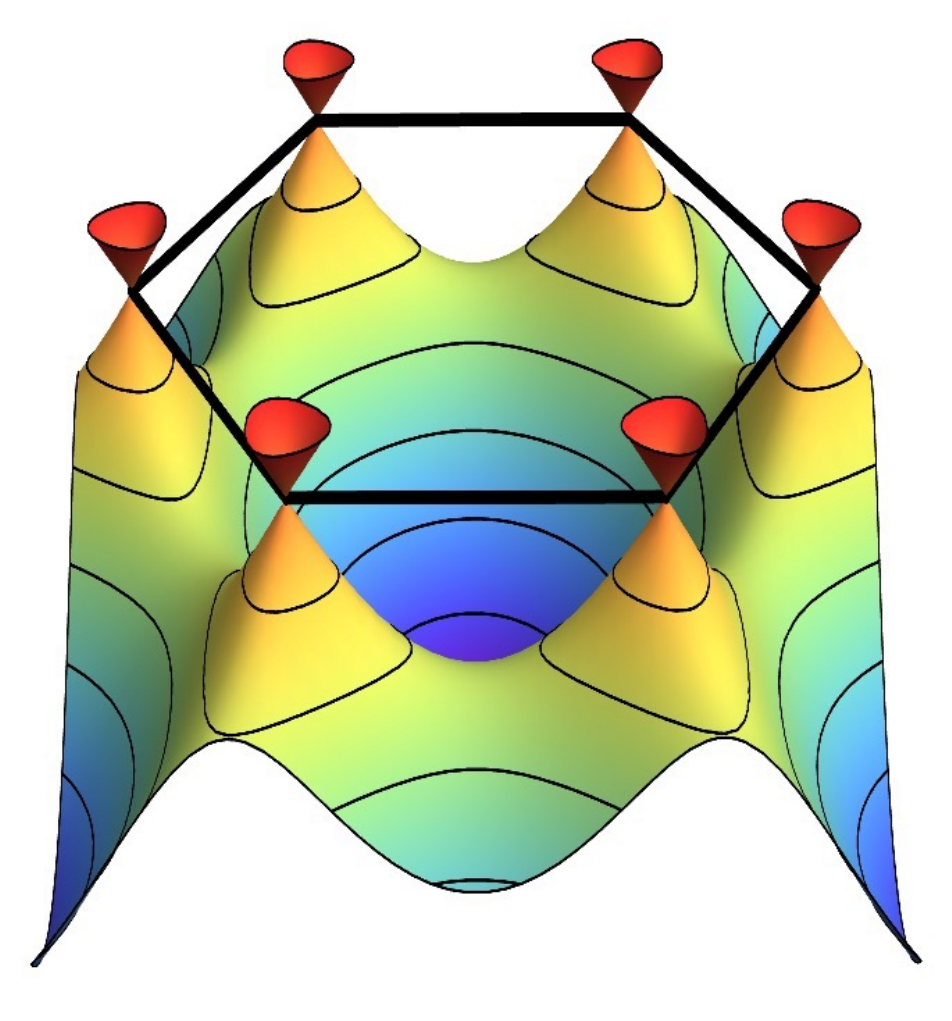}
  \includegraphics[width=0.54\linewidth]{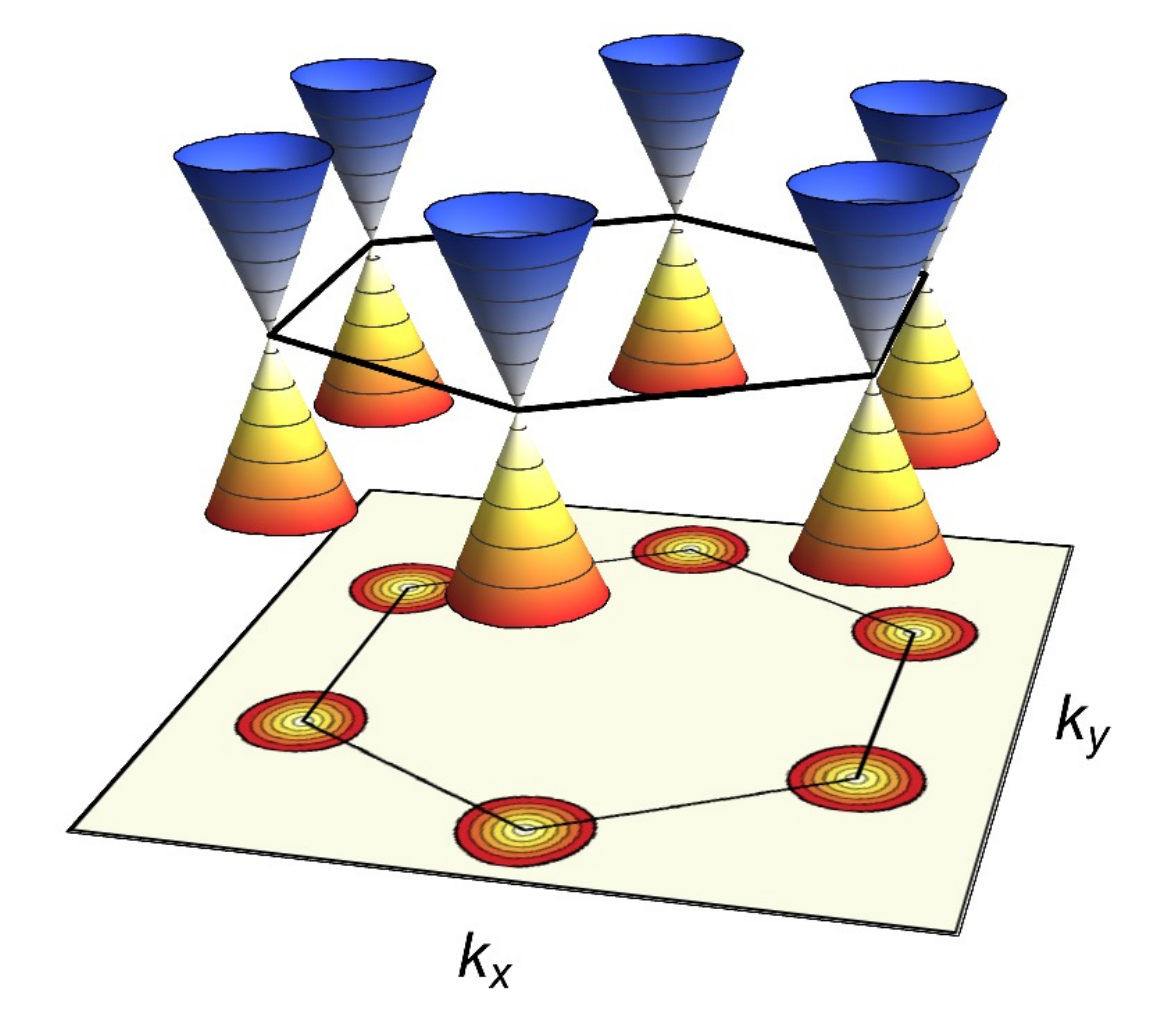}
  \caption{Left: Excerpt of the dispersion relation $E(\v k)$ with six conical Dirac points $\K^{(s)}$ (red) forming a regular hexagon (black).
  Right: The same cones magnified.
  \label{fig:G-Spectrum}}
\end{figure}

In regular graphene, where all $t_{\v n,l} = t_0$, the spectrum of the Hamiltonian exhibits conical Dirac points $\v K^{(s)}$ at which the dispersion relation is approximately linear $E(\k) \sim |\k-\K^{(s)}|$,
where $\K^{(0)} = \left( 0, {4\pi}/{(3\sqrt{3} d_0)} \right)$ and $K^{(s)}$ are obtained by rotation of $\v K_0$ by an angle $\frac{\pi}{3} s$ with $s=0,...,5$, as shown in Fig \ref{fig:G-Spectrum} (cf. \cite{Katsnelson-GrapheneBook} for more fundamentals of graphene theory).
Since only two of them are physically inequivalent we will choose the pair $\K^{(0)}, \K^{(3)}$ and denote it $\K^\pm = \left( 0, \pm{4\pi}/{(3\sqrt{3} d_0)} \right)$.

In the long wave limit, the discrete tight--binding Hamiltonian can be approximated by the two-dimensional Dirac Hamiltonian \cite{CastroNeto+Guinea+Novoselov+Geim-GrapheneReview}
\begin{equation} \label{eq:Ham-Dirac}
   H_D =  i v_F \sigma^l \left(\p_l - i K^{(s)}_l \right)
\end{equation}
separately for each of the Dirac points $s$.
Here, $v_F=\frac{3}{2} t_0 d_0$ is the Fermi velocity of the excited electrons and $\sigma^l$ are Pauli matrices.
As long as regular planar graphene is considered, the constant $\v K^{(s)}$ vectors can be gauged away by multiplication of the wavefunction with an appropriate phase.

\begin{figure}
  \includegraphics[width=0.65\linewidth]{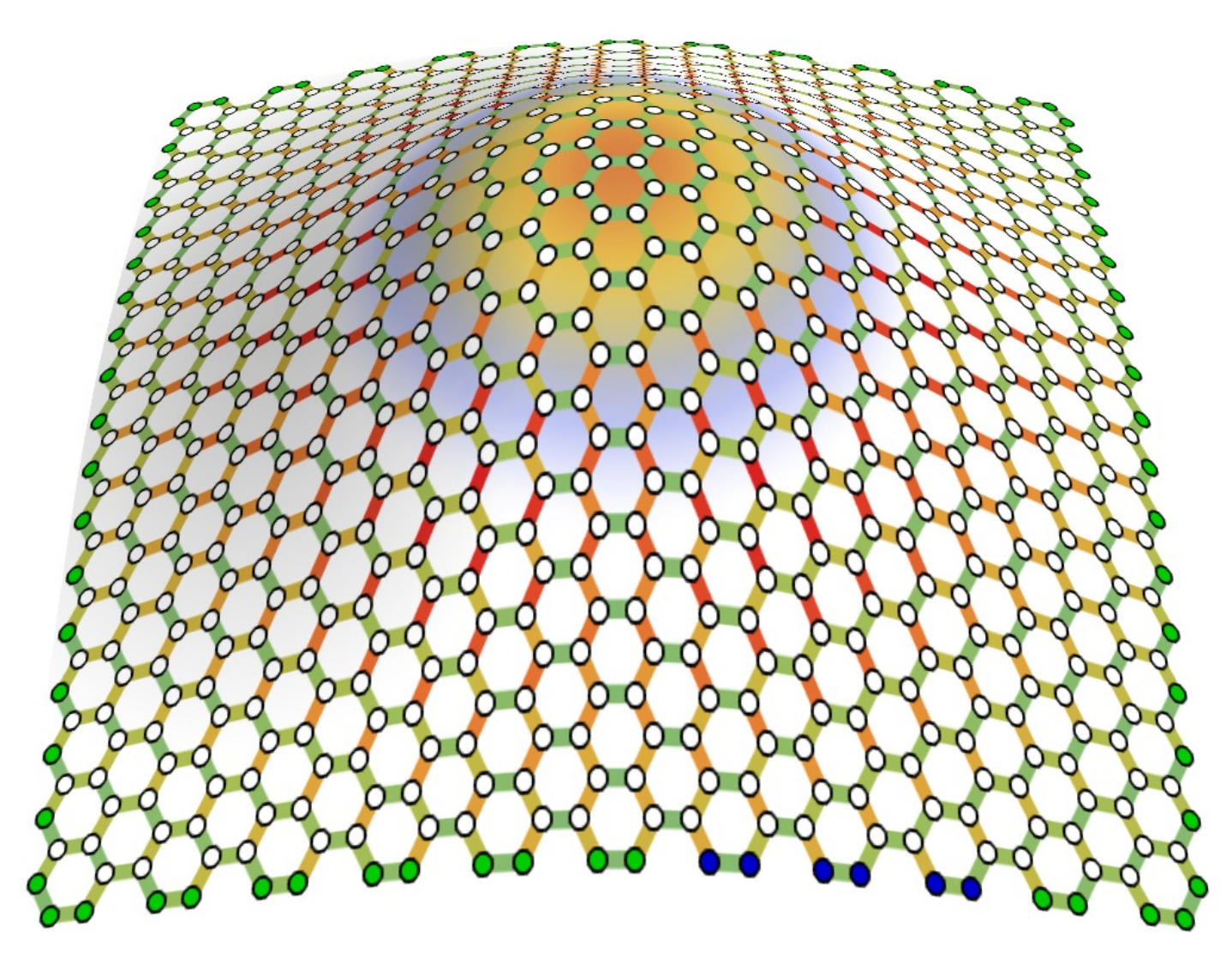}\\
  \includegraphics[width=0.65\linewidth]{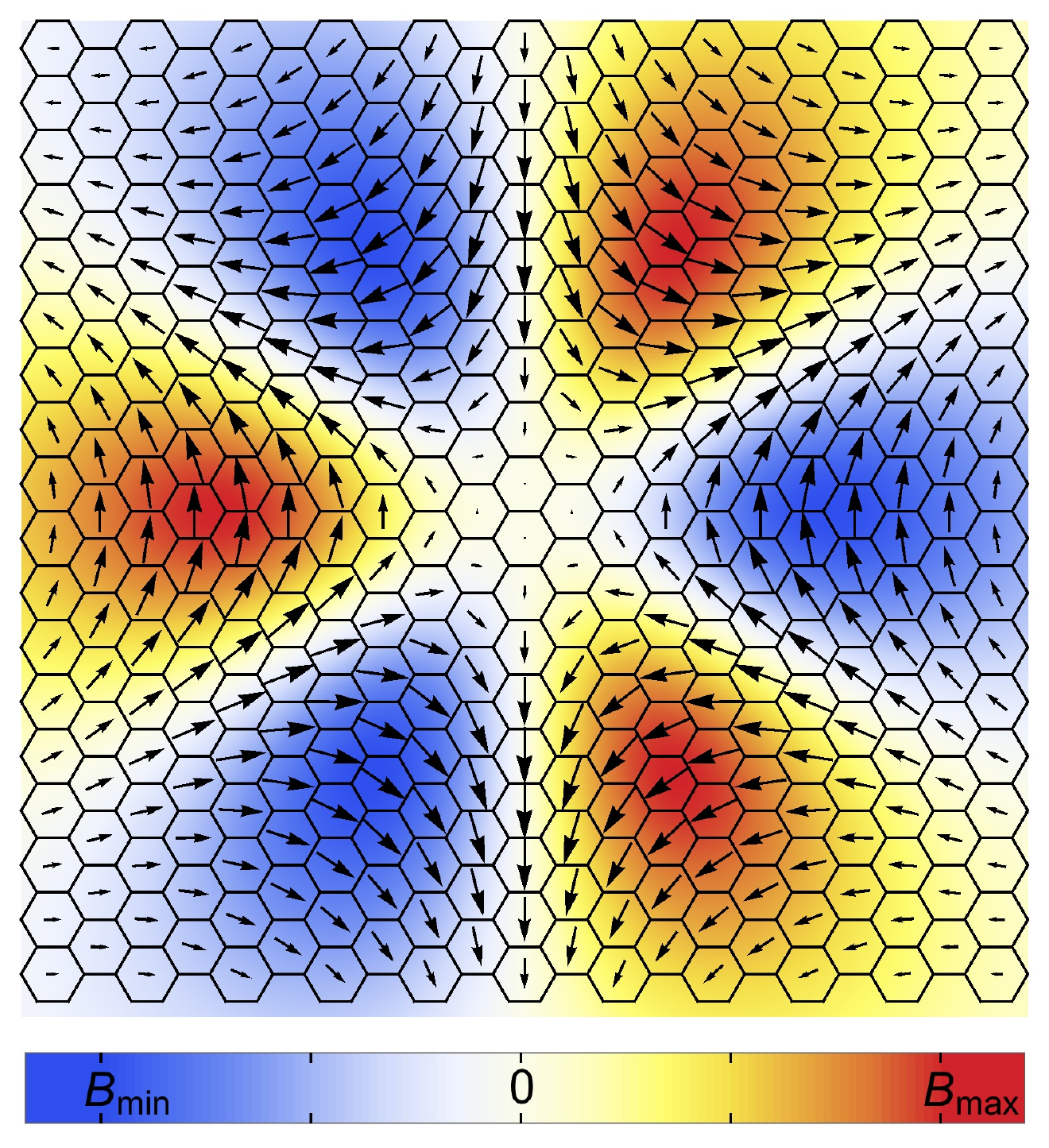}
  \caption{Locally deformed graphene. 
  Top: Modified hopping parameters (colors of the links) and Gauss curvature (background color shading).
  Bottom: Cell--averaged pseudo--magnetic vector potential (arrows) and pseudo--magnetic field $\t B$ (background color shading).
  \label{fig:Struct+Ricci+A+B}}
\end{figure}

In this work we concentrate on small out-of-plane perturbations described by the height function $h(x,y)$ (cf. Fig \ref{fig:Struct+Ricci+A+B}, top). We assume for simplicity that the carbon atoms are just lifted by the deformed surface in the perpendicular direction $z_{\v n} = h(x_{\v n},y_{\v n})$ while keeping their original $(x_{\v n},y_{\v n})$ coordinates in the plane.
This approximation is acceptable as long as the perturbations of the distances stay small%
\footnote{Otherwise the true positions of the atoms should be calculated from first principles \cite{Katsnelson-GrapheneBumpMidgapState+Relaxation}, e.g. via a DFT method, similarly to  the shape of the surface itself which will be chosen by the graphene atoms under the action of a perpendicular force.}, i.e. $\d \v d_{\v n,l} = \leps_{\v n}\,  \v d_{\v n,l} \ll d_0$, where $\leps$ is the strain tensor describing deformation of the graphene lattice \cite{Pereira+CastroNeto-UniaxialStrain}.
Knowing the positions of the atoms the perturbations of their distances can be directly calculated via the Euclidean formula and for small deformations approximated by
$ \d |\v d_{\v n,l}| \approx (\d z_{\v n}/d_0)^2/2 \approx (\p_l h)^2/2$
where $\p_l h$ is a directional derivative along the link $\da$.

Accordingly, the hopping parameters change slightly to become
$\t t_{\v n,l}=t_0+\d t_{\v n,l}$ and vary slowly over the lattice.
Applying an empirical relation between the distances $\v d_{\v n,l}$ and the hopping terms $\t t_{\v n,l}$ we get%
\footnote{We ignore also the modification of hopping parameters $\t t_{\v n,l}$ due to non--orthogonality of $\pi$--orbitals and rehybridization \cite{Graphene-curvature-hopping} since this effect is rather small in our case.}
\begin{equation} \label{eq:mod-hopping}
  \t t_{\v n,l} = t_{0} \exp(-\beta\, \d | \v d_{\v n,l}|/{d_{0}})
\end{equation}
with $\beta=3.37 $ \cite{CastroNeto-Graphene+Strain-hopping, Pereira+CastroNeto-UniaxialStrain, Bump+Transport}.
For small length variations $\d | \v d_{\v n,l}| \ll d_0$ the latter can be approximated by the linear relation
$ {\d t_{\v n,l}}/{t_{0}} = -\beta\, {\d |\v d_{\v n,l}|}/{d_{0}}$ which will allow us to relate $\d t_{\v n,l}$ linearly to the strain tensor $\leps_{\v n}$.

It can be shown that the Hamiltonian \eqref{eq:Ham-hopping} with such modified hopping parameters%
\footnote{Only perturbations of the hopping parameters $t_{\v n,l}$ can have physical consequences. Perturbations of the link vectors $\v d_{\v n,l}$, taken into account e.g. in \cite{Pereira-StrainedGraphene, Moldovan+Peeters-GrapheneStrainRevisited}, can modify the shape of the Brillouin zone and the relative positions of the $\K^{(s)}$ points but can only have a pure gauge character.}
leads in the long wavelength limit to the continuous Hamiltonian \cite{Juan+Vozmediano-SpaceDepFermiVelG, Juan+Manes+Vozmadiano-GaugeFieldsFromStrain, Naumis-NonuniformlyStrainedGraphene, NS-GrapheneElasticDef}
\begin{multline}
   \t{\op{H}} = \int d^2x \sqrt{|\t g(\v x)|}\, \Psi^\dagger(\v x)\, i v_F\, \sigma^a\, \t e_a^{\;l}(\v x)\cdot \\
   \cdot\left(\p_l - i \t K^{(s)}_l(\v x) + \t\Gamma_l(\v x)\right) \Psi(\v x)
\end{multline}
resembling the one for the Dirac fermions in curved space.
The low energy electronic excitations would then satisfy an effective%
\footnote{All effective fields will be distinguished by a tilde.}
two-dimensional Dirac equation $i\p_t \Psi = \t H_D \Psi$ with the evolution generator given by
\begin{equation} \label{eq:Ham-Dirac-deformed}
   \t H_D = i v_F \sigma^a \t e_a^{\;l}(\v x) \left(\p_l - i \t K^{(s)}_l(\v x) + \t\Gamma_l(\v x) \right).
\end{equation}
Here, $\te{a}(\v x)$ ($a=1,2$) plays the role of a local frame (\textit{zweibein})
and gives rise to an effective (inverse) metric 
\begin{equation}
   \t{g}^{ij}(\v x) = \t e_a^i(\v x)\ \t e_b^j(\v x)\ \d^{ab}
\end{equation}
describing an emergent curved geometry in which the electronic excitations propagate.
$\t K^{(s)}_l(\v x)$ is a vector potential whose curl gives rise to an effective pseudo--magnetic field
\begin{equation}
  \t B^{(s)}(\v x) = \mathop{\rm rot} \tK{}^{(s)}(\v x) = (-1)^s \t B(\v x)
\end{equation}
which in two dimensions has only one component. 
$\t B^{(s)}$ acts oppositely on the excitations in different valleys $s$ (hence the prefix ``pseudo'')
and has therefore to be distinguished from a true magnetic field which would break the time--reversal symmetry of the system. 
The latter situation is excluded here because the perturbations of the hopping parameters in \eqref{eq:Ham-hopping} are purely real.
$\t\Gamma_l(\v x)$ corresponds to the spin--connection and guarantees hermiticity of the above Hamiltonian when the frame $\t e_a^i(\v x)$ is position--dependent \cite{Juan+Vozmediano-SpaceDepFermiVelG, Juan+Manes+Vozmadiano-GaugeFieldsFromStrain}. 
In contrast to the vector potential and the frame perturbations, both proportional to the strain, spin--connection is 
proportional to its gradient.
It has no classical counterpart but in quantum scattering on steep surface deformations containing closed geodesics it can contribute to transmission resonances at wavelengths comparable with the size of the deformation \cite{GeodesicScatteringInTopInsulator, Guinea+Tagliacozzo-SpinConInTopInsulator}.
For shallow deformations varying smoothly over the surface, such as considered here, the spin--connection becomes subdominant and can be skipped. 

Both remaining fields are continuous extrapolations of their discrete counterparts on the lattice, given in Appendix \ref{App:fields-from-hopping}, and their values can be related to an abstract strain tensor $\teps$ which transforms the local frame%
\footnote{There is a small discrepancy between the frame transformation given in \eqref{eq:local-frame} and $\t e_a^i$ interpreted as local Fermi velocity in \cite{Juan+Vozmediano-SpaceDepFermiVelG, Juan+Manes+Vozmadiano-GaugeFieldsFromStrain} 
by a strain--trace term ${\t\eps}^{\ i}_i$. It is related to different normalizations of the wavefunction $\Psi(\v x)$. 
Since on the lattice $\sum_{\v n} \bar\psi_{\v n}\, \psi_{\v n} = 1$ in curved space $\int \bar\Psi\, \Psi\, \sqrt{g}\, d^2x = 1$ 
which forces us to choose the discretization rule $\psi_{\v n} = g^{1/4}(\v x_{\v n}) \Psi(\v x_{\v n})$ here.
}
\begin{equation} \label{eq:local-frame}
  \te{a}(\v x) = (\1 - \teps(\v x))\, \e{a},
\end{equation}
defines the metric of the curved continuous space
\begin{equation}
  \t{g}^{ij}(\v x) = \d_{ij} - 2\,{\t\eps}_{ij}(\v x)
\end{equation}
and the pseudo--magnetic vector potential
\begin{equation}
  \tK^{(s)}(\v x) = \K^{(s)} + (-1)^s \frac{\beta}{2} 
  \begin{pmatrix}
    -2 \t\eps_{xy},& \t\eps_{yy}-\t\eps_{xx}
  \end{pmatrix}
\end{equation}
\cite{VozKatsGui-review,CastroNeto+Guinea+Novoselov+Geim-GrapheneReview, Pereira+CastroNeto-UniaxialStrain}.
It is not a priori known what the effective geometry will look like since, for the given curved surface of graphene, the effective metric  is obtained by the expansion of the Hamiltonian \eqref{eq:Ham-hopping} around the Dirac points shifted by $\tK^{(s)}$ \cite{Volovik-EmergentGeometryGraphene}.
However, due to the assumed proportionality of $\d t_{\v n,l} \sim \beta\,\d |\v d_{\v n,l}|$, the effective strain tensor $\teps$ is proportional to the real strain applied to graphene, $\teps = \beta\,\leps$, and thus the effective geometry for the electronic excitations is nearly identical to the real geometry of the deformed graphene sheet but the deformation is magnified by the factor $\beta>1$.

\begin{figure}
  \includegraphics[height=5cm]{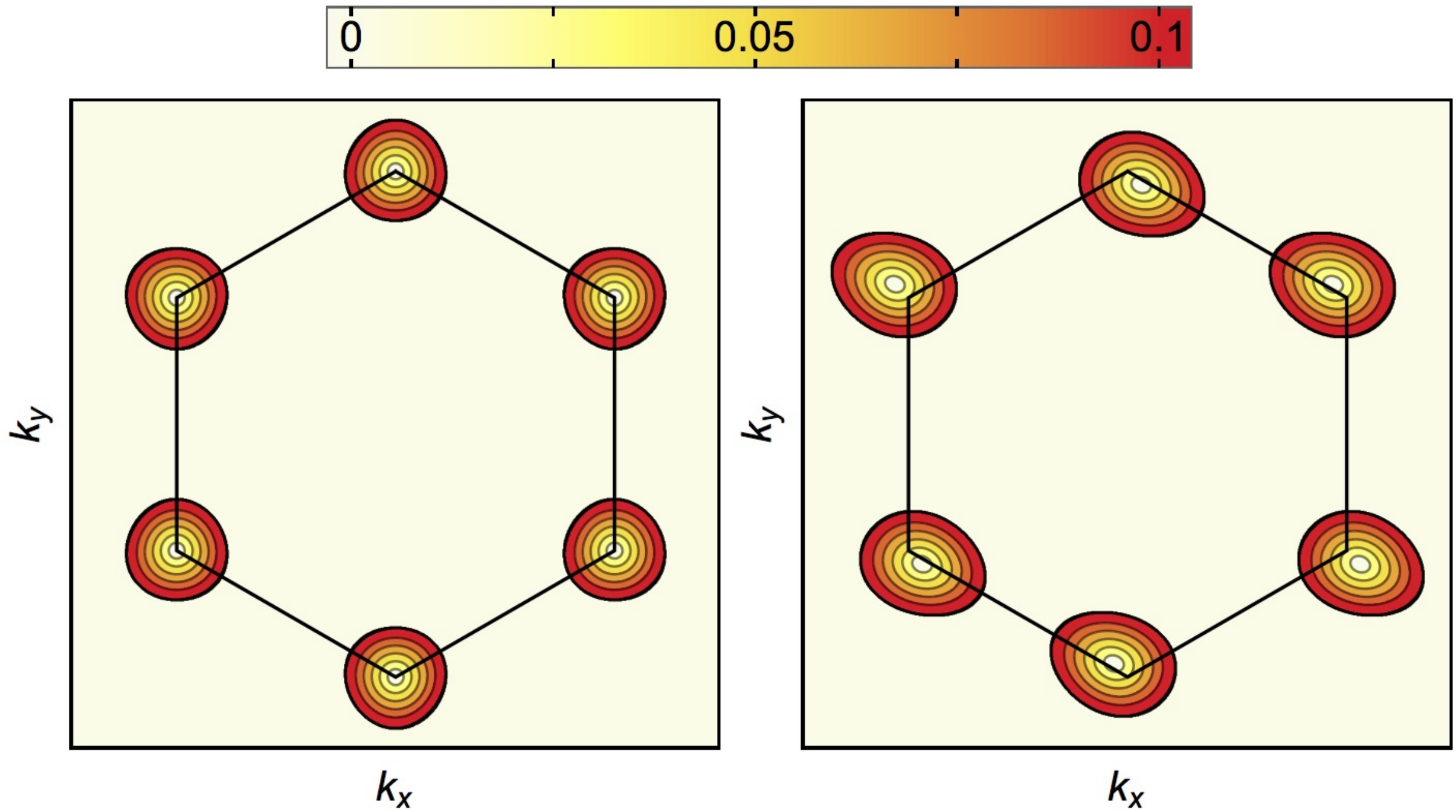}
  \caption{
  Non-deformed (left) and deformed (right) cones in a top-projection.
  Energy scale in units of $t_0$.
  \label{fig:G-Spectrum-pert}}
\end{figure}

When the graphene lattice is perturbed the Dirac cones do not have a global meaning any more.
Instead, everywhere on the lattice local cones can be associated to the Dirac operator in \eqref{eq:Ham-Dirac-deformed} along which wavefronts of local perturbations would propagate (cf. microlocal analysis of operators \cite{Baer+Fredenhagen-QFT-curvedST}).
While the pseudo--magnetic vector potential $\tK{}^{(s)}(\v x)$ locally shifts the Dirac cones, the effective inverse metric locally deforms them in the momentum space (see Fig. \ref{fig:G-Spectrum-pert}). In consequence, the pseudo-momentum vectors $\v k$ belonging to the deformed cones located at the shifted Dirac points satisfy locally
\begin{equation}
  \t{g}^{ij}(\v x) \left(k_i - \t K_i{}^{(s)}(\v x)\right) \left(k_j - \t K_j{}^{(s)}(\v x) \right) = 0.
\end{equation}

In the continuous approximation the fields can be written in terms of the height function $h(\v x)$ and its derivatives:
in the lowest order the effective strain reads
\begin{equation}
  \t\eps_{ij}(\v x)=\frac{\beta}{2}\,\p_i h(\v x)\, \p_j h(\v x)
\end{equation}
and the pseudo--magnetic field $\t B(\v x) = \mathop{\rm rot} \tK(\v x)$ is
\begin{equation}
  \t B(\v x) =  \p_x\, \t\eps_{yy}(\v x) - \p_x\, \t\eps_{xx}(\v x)  + 2\,\p_y\, \t\eps_{xy}(\v x).
\end{equation}

In the examples discussed below we will consider rotationally symmetric elastic deformations of graphene for which the height function $ h(x,y) = h(r)$ uniquely parameterizes the lattice geometry.
Such geometries seem to be well under experimental control as they appear when a localized perpendicular force is applied to the surface, e.g. from an STM--head \cite{DefGraphene-STM-tip}, and forms a tip--like deformation.
Then, the last formula simplifies (in polar coordinates) to
\begin{equation} \label{eq:pseudomagneticF}
  \t B(r,\varphi) = \frac{\beta\, \cos(3 \varphi)}{2}\, h'(r) \left[ h''(r) - \frac{1}{r} h'(r) \right].
\end{equation}
The angular prefactor $\cos(3 \varphi)$ changes sign six times on the interval $(0,2\pi)$ and hence creates 6 zones of alternating sign of $\t B$ as can be seen in Fig. \ref{fig:Struct+Ricci+A+B}.

The definitions of both emergent fields, the metric and the pseudo--magnetic field, directly in terms of the change of the hopping parameters $\d t_{\v n,l}$ are given in Appendix \ref{App:fields-from-hopping}. 
While here, $\d t_{\v n,l}$ arise in response to an elastic deformation of the graphene structure, the developed formalism can be applied to any lattice system described by the Hamiltonian \eqref{eq:Ham-hopping} in which $\d t_{\v n,l}$ can be of a different origin (cf. hopping of atoms in optical lattices \cite{Lewenstein-OptLatCurvature, Minar-CurvedSTinOptLat, NS-CurvedSpaceInLab}).

\subsection{Geodesic lines in curved continuous geometry}

In order to develop an efficient way of predicting the current flow paths in deformed graphene we will make use of the geometrical optics (eikonal approximation) in which the propagation of waves is replaced by the tracing of rays. The latter satisfy ``equations of motion'' typical for point-like particles.
Eikonal approximation applied to the Dirac equation in curved space leads to the Mathisson--Papapetrou equation of motion describing a spinning particle in curved space \cite{Mathisson-Papapetrou-from-Dirac}.

Since the issue of spin in graphene is quite delicate we postpone its treatment to future work and ignore the spin degree of freedom here\footnote{Here, with spin we mean the pseudo--spin appearing effectively due to the symmetries of the honeycomb lattice \cite{Graphene-Spinors}. The real spin of electrons is usually ignored in graphene anyway.}.
Then, by squaring the Dirac Hamiltonian and applying the eikonal approximation, we arrive at the geodesic equation for null geodesics (due to massless fermions) in a given curved 2D--surface.
Note that in a static, i.e. time-independent, 2+1D geometry the full spacetime 2+1D geodesics match with the shortest (extreme) lines of the spatial 2D geometry.
Therefore, it is sufficient to solve the geodesic equation for the 2D curved surface
\begin{equation} \label{eq:geodesic}
   \frac{dv^i}{d\tau} + \t\Gamma^i_{kl} v^k v^l = \sqrt{\t g}\ \t g^{ij} \epsilon_{jk}\, v^k \t B^{(s)}
\end{equation}
where the ``velocity'' $v^i(\tau) = dx^i(\tau)/d\tau$. 
Here
$\t\Gamma^i_{kl} = \frac{1}{2}\, \t g^{ij}(\p_k\, \t g_{jl} + \p_l\, \t g_{kj} - \p_j\, \t g_{kl} )$
denote the Christoffel symbols for the metric $\t g_{ij}$ and $\epsilon_{ij}$ is the completely anti--symmetric Levi--Civita symbol ($\epsilon_{12}=-\epsilon_{21}=1$).
We are interested in a family of geodesics starting parallel to each other and representing a current injected at the contact with the initial momentum $p_i = k_i - \t K_j{}^{(s)}$, with $\v k$ being the wave vector satisfying $v_F|\v k| =  E$.
Therefore, the initial velocity $v^i(\v x(\tau_0))$ for a geodesic starting at the point $\v x(\tau_0)$ should be chosen as
\begin{equation} \label{eq:initval-geodesic}
  v^i(\v x(\tau_0)) = \t g^{ij}(\v x(\tau_0))\, \left(k_j -  \t K_j{}^{(s)}(\v x(\tau_0)) \right),
\end{equation}
i.e. different for each geodesic line. 

If $\t B^{(s)}$ is negligibly small the solutions do not depend on the value of the initial velocity (i.e. depend only on its direction) nor on the initial energy (also not on  its sign) and hence particles and antiparticles follow the same lines.
The presence of the magnetic field $\t B^{(s)}$ breaks this symmetry and deflects particles and antiparticles in opposite directions. Also, the initial value of the velocity (energy) starts to play a role -- the lower it is the more the magnetic field has influence on the trajectory.

\begin{figure}[ht]
  \includegraphics[width=0.75\linewidth]{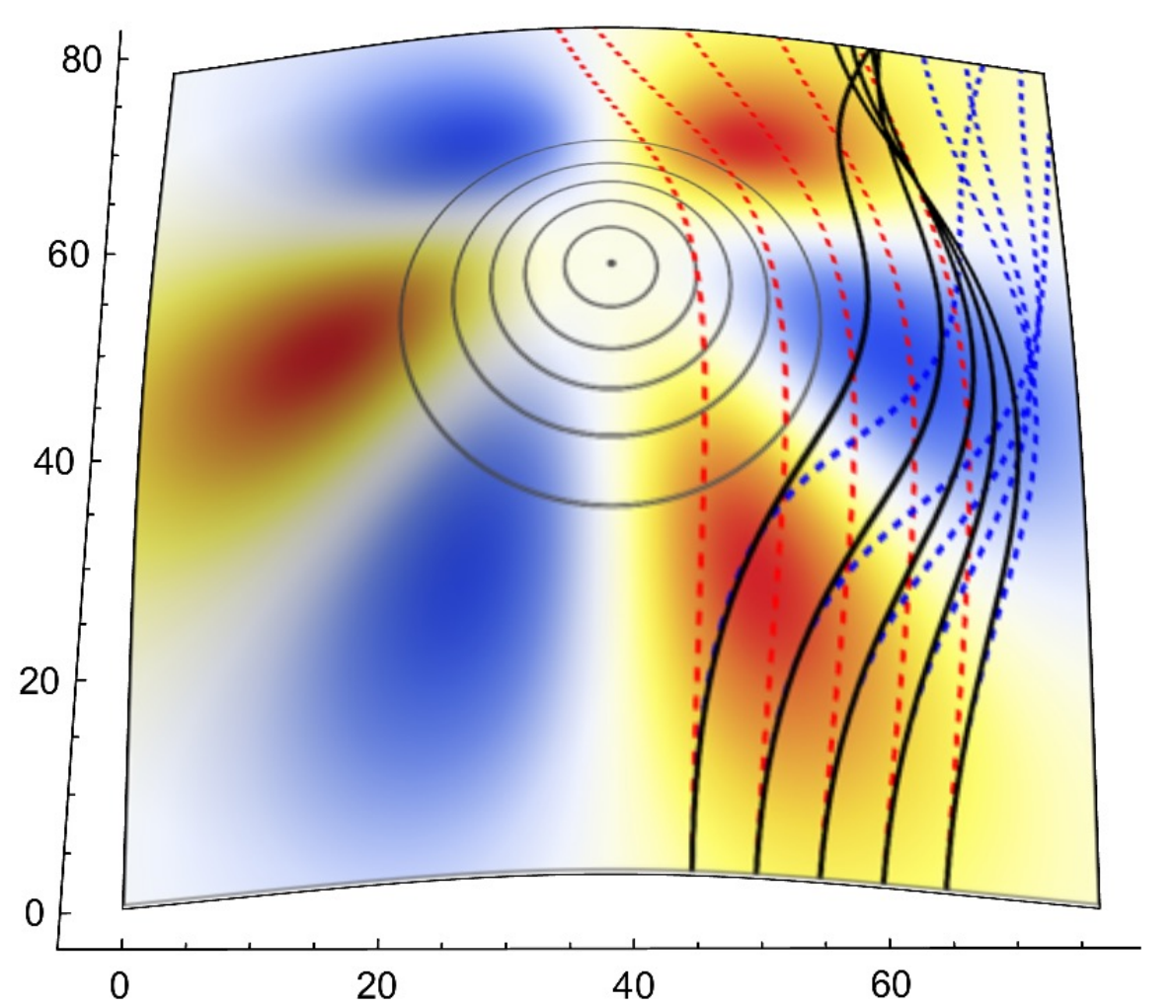}
  \caption{Comparison of trajectories calculated for contributions from curved geometry (red dashed lines), magnetic field (blue dashed lines) and both (black solid lines). Background color represents the magnetic field $\t B$ while concentric circles are isolines of the Gaussian curvature.
  \label{fig:Geo-GB0}}
\end{figure}

The curvature of a typical bump is clearly positive in the middle (and only slightly negative outside, cf. Fig. \ref{fig:Struct+Ricci+A+B}, top) and hence it bends the geodesic lines on the surface towards the center -- as attractive gravitational forces do.
The pseudo--magnetic field, due to its six-fold symmetry (cf. Fig. \ref{fig:Struct+Ricci+A+B}, bottom), bends the geodesics inwards and outwards in an alternating way.
The two forces, the geometric $\t\Gamma^i_{kl}$ and pseudo--magnetic $\t B$, are both proportional to the gradient of the strain tensor $\varepsilon_{ij}$ and hence to the product of first and second derivatives of the height function, symbolically $\sim \p h \cdot \p^2 h$.
Hence, both contributions are of the same order of magnitude so the correct prediction of electric currents in deformed graphene will require taking both of them into account. 
These two contributions to the bending of trajectories are compared in Fig. \ref{fig:Geo-GB0} for a typical bump geometry.

\section{Transport in deformed graphene} 
\subsection{The NEGF method}
\label{sec:NEGFmethod}

\newcommand{\fig}[1]{figure~\ref{fig:#1}}
\renewcommand{\Im}[1]{\ensuremath{\mathrm{Im} \left(#1\right)}}
\newcommand{\I}{\mathrm{i}}
\newcommand{\Sg}{\Sigma}
\newcommand{\dg}{\dagger}
\newcommand{\Tr}{\mathrm{Tr}}

In order to determine the current flow paths in deformed graphene we calculate numerically the local current density $I_{\v n,\v m}$ between the neighboring lattice sites $\v n, \v m$ and plot its integral lines.
To find the current in the presence of a stationary source, injecting electrons into the ribbon at a constant pace, we apply the non--equilibrium Green's function method (NEGF).
As this method has been described in various textbooks, see
e.g. \cite{Datta1997, Datta2005, NEGF-Formalism}, here we only summarize briefly the required equations.

Starting from the tight--binding Hamiltonian $\op{H}$ of a deformed graphene ribbon \eqref{eq:Ham-hopping}, the matrix elements of the Green's function $\op{G}$ are given by
\begin{equation} \label{eq:GreensF}
  G_{\v n,\v m} = \<\psi_{\v n} | (\op{H}-E\,\1-\op{\Sg})^{-1} | \psi_{\v m}\>.
\end{equation}
$E$ is the energy of the injected electrons and
\begin{equation} \label{eq:Self-Energy}
  \op{\Sg} = -\I \eta \sum_{\v n \in \lEdge} \ket{\psi_{\v n}}\bra{\psi_{\v n}},
\end{equation}
is an imaginary self--energy by means of which
we introduce absorbing boundaries \cite{Stegmann2013}, mimicking infinite dimensions of the graphene sheet.
(The sum runs over the edge atoms $\lEdge$, cf. green sites in Fig. \ref{fig:Struct+Ricci+A+B}, and $\eta>0$ is a constant.)
These boundaries absorb impinging particles and suppress finite system size effects, such as standing waves between the boundaries of the system.

The electrons are injected in the graphene ribbon by a contact $\lCon$ (blue sites in Fig. \ref{fig:Struct+Ricci+A+B}, top), which we model by the inscattering function
\begin{equation} \label{inscattering-function}
  \op{\Sg}^{\text{in}}_\lCon = \nu \sum_{\v n,\v m \in \lCon} \chi_{\v n}^*\, \chi_{\v m} \ket{\psi_{\v n}}\bra{\psi_{\v m}},
\end{equation}
where
$\nu>0$ is a constant and
$\chi_{\v n}$ are amplitudes encoding the initial energy and momentum of the injected plane wave at the contact $\lCon$.
They are given by separate expressions on both sublattices $\lA, \lB$, namely
\begin{equation} \label{phase-factors}
  \chi_{\v n} = \left\{ \begin{array}{ll}
    C_{-} e^{i(\v k+\v K^-)\v n} + C_{+} e^{i(\v k+\v K^+)\v n},  & \v n \in \lA \\
    \sigma \, C_{-} e^{i(\v k+\v K^{-})\v n + i \phi} - \sigma \, C_{+} e^{i(\v k+\v K^{+})\v n - i \phi}, & \v n \in \lB
  \end{array} \right.
\end{equation}
where $\phi=\arctan(k_x/k_y)$, $\sigma = \text{sign}(E)$ and $C_{\pm}$ are amplitudes of the excitations around the $\K^{\pm}$ valleys.
This inscattering function corresponds to the injection of plane waves with momentum $\v k$ and energy $E$. 
Since we consider idealized contacts the injection of electrons 
does not affect their propagation in the graphene ribbon 
and thus we take $\op{\Sg}_\lCon=0$ for the self-energy of the contact $\lCon$.

The local current of electrons, which originate from the contact $\lCon$ with energy $E$ and which flow from
atom $\v n$ to the neighboring atom $\v m$, is given by \cite{Caroli1971, Cresti2003}
\begin{equation} \label{eq:current-NEGF}
  I^\lCon_{\v n,\v m}= \Im{t_{\v n,\v m}^*\, {\mathcal G}^\lCon_{\v n, \v m}},
\end{equation}
(in the natural units $e\, t_0/h$) where
\begin{equation} \label{eq:CorrelationF}
  \bm{\mathcal{G}}^\lCon = \op{G}\, \op{\Sg}^{\text{in}}_\lCon\, \op{G}^\dg.
\end{equation}
is the correlation function of the injected electrons.
Interestingly, the lattice current formula \eqref{eq:current-NEGF} can also be derived as discretization of the Dirac current in curved space, see Appendix \ref{App:discrete-Dirac-current}.

Since, in general situations, we deal in graphene with contributions from both inequivalent Dirac points $\v K^{(\pm)}$, interference effects in the current (due to its quadratic dependence on the wavefunctions) are expected.
In order to eliminate the highly oscillatory behavior on the lattice scale from the plots, we average the numerical values of the current over the hexagons.
We leave the examination of the interference patterns to a future work.


\subsection{System parameters}
\label{sec:SystemParameters}

We consider rectangular graphene ribbons of size varying from $80 \times 50$ to $240 \times 180$ carbon rings in the armchair and zigzag direction, respectively. This corresponds to sizes $L_x \times L_y$ between $120 \times 85$ and $360 \times 310$ (multiples of $d_0$). The bump--like deformation (cf. Fig \ref{fig:Struct+Ricci+A+B}, top)
\begin{equation}
  h(r) = \frac{h_0}{1+(r/r_0)^2}
\end{equation}
is placed in the center of the ribbon with a spatial extension $r_0$ varying between $100$ and $200$ and the height $h_0$ varying between $0.5 r_0$ and $1.5 r_0$. The ratio $h_0/r_0$ controls the steepness of the bump and thus the maximal strain which should not exceed 10\% for the continuous model to hold \cite{Guinea+Katsnelson+Geim-LandauLevelsInStrainedG}.

We choose the electric current to flow along the zigzag direction, from a contact placed on the armchair edge.
Its width $D$ is adjusted to the wavelength $\la$ of the injected quantum current (cf. discussion below) and takes values between $10\%$ and $30\%$ of the edge length $L_x$.

\begin{figure}[ht]
  \includegraphics[width=0.8\linewidth]{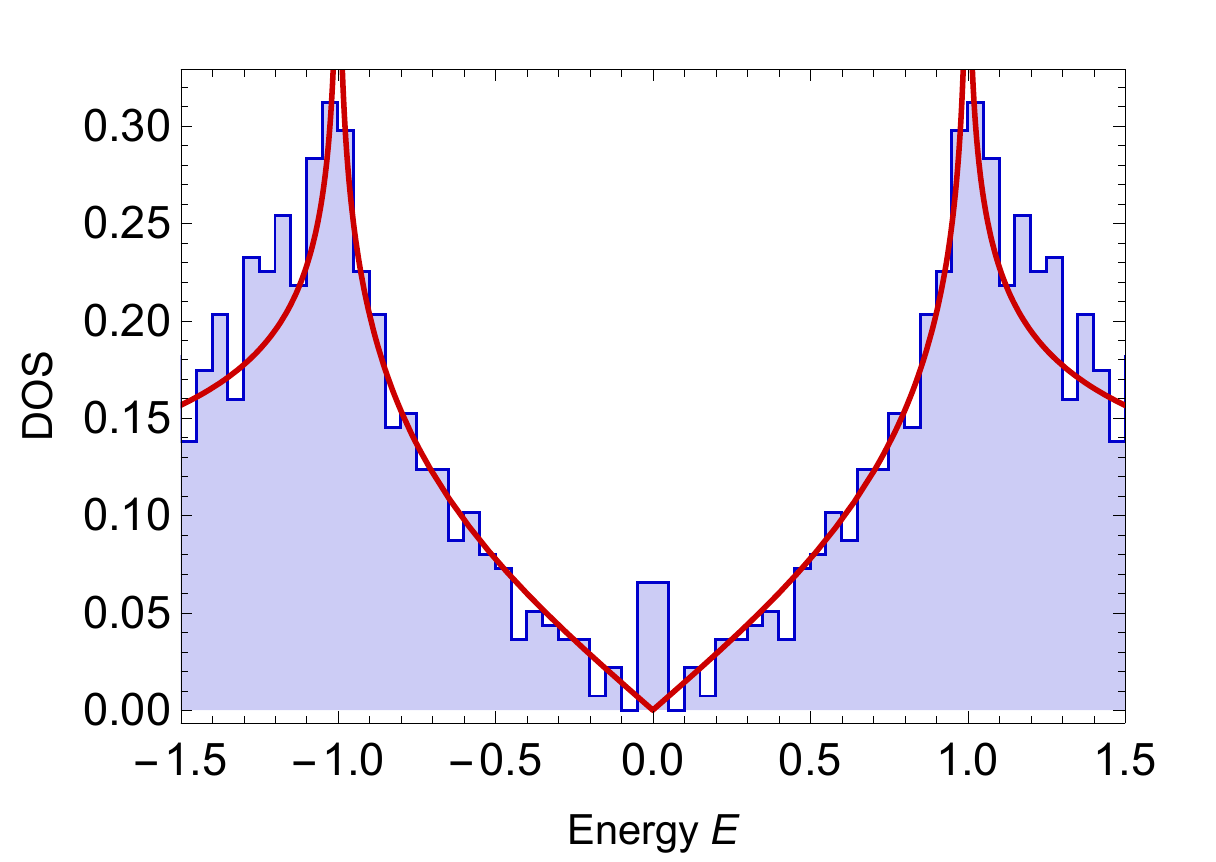}
  \caption{Density of states (DOS) in a flat graphene ribbon of size $60\,d_0 \times 50\,d_0$ (blue histogram)
  compared to that of an infinite planar graphene (red line).
  The small ribbon size has been chosen to emphasize the discrepancies.
  The peak at $E=0$ corresponds to edge states which are typical for finite size nanoribbons \cite{NanoribbonProperties, NanotubesProperties}
  and should not be confused with midgap states which appear first for very high bumps with $h_0^2\geq r_0$ \cite{Guinea+Katsnelson+Vozmediano-GrapheneBumpMidgapState}.
  \label{fig:DOS}}
\end{figure}

In our numerical simulations we can consider only graphene sheets of finite size (nanoribbons).
Consequently, the available energy spectrum is discrete.
However, for large ribbons the separation of the discrete energies shrinks to zero and reaches the continuous band structure in the limit.
To test the granularity, it is instructive to plot the density of states and check the behavior near $E=0$ which is crucial for our approximations (cf. Fig \ref{fig:DOS}).
A simple estimation of the length scales leads to the gap%
\footnote{The isolated states at $E=0$ visible in Fig. \ref{fig:DOS} correspond to dispersionless zz--edge states present in graphene nanoribbons which are known to not contribute to the electronic transport \cite{CastroNeto+Guinea+Novoselov+Geim-GrapheneReview, Dresselhaus-GrapheneEdgeStates}.}
$\Delta E = 3\pi/L$ where $L = \min(L_x,L_y)$.
It reflects the absence of long waves with wavelengths $\la$ larger than the system size $L$ and limits from below the range of admissible current energies $E$.
In our systems $\Delta E$ is between $0.03$ and $0.11$ and is kept well separated from the considered current energies.

On the one hand, for small energies $E$ the wavelengths $\la \approx \frac{3\pi}{E}$ must satisfy $\la \ll L_x, L_y$
in order to be compatible with the size of the ribbon. On the other hand, they must be much larger than the interactomic distance $\la \gg d_0$ for the continuous approximation to hold.
Further on, for the geometrical optics approximation to hold, the wavelengths $\la$ must be shorter than the scale of the geometric structures on which the wave is supposed to scatter which here means $\la \ll r_0$.
This gives a hierarchy of length scales $d_0 \ll \lb \ll r_0 < L_x, \, L_y$ which must be satisfied in all numerical computations.

\subsection{Propagation of plane waves}

In order to see the electric current flowing along the geodesics in the curved geometry of graphene it is necessary to stay close to the regime of plane waves propagating through the ribbon.
While exact plane waves would require infinitely wide regions, in finite graphene ribbons we will have to deal with disturbing boundary effects. 
As a solution, we choose a finite contact, in the middle of one boundary and well separated from others, at which the wave will be injected locally as plane.
In a sense, it corresponds to a single-slit diffraction:
a hypothetical plane wave comes from outside the ribbon and entering the ribbon gets diffracted at the slit (here contact).
For narrow contacts the wave will naturally spread across the ribbon.
Contacts which are wider than the wavelength $\lambda$ produce interference effects at angles $\sin(\theta) = \lambda / D$ where $D$ is the contact width.
Hence, at the opposite end of the ribbon the width of the ``beam'' (measured between the two first interference minima) is $D' \approx L \sin(\theta) = L \lambda / D$, where $L$ is the length of the ribbon.
Since we want the current flow to have approximately constant width during its propagation we choose the parameters so that $D \approx D' \approx \sqrt{L \lambda}$ holds.

\begin{figure}[ht]
  \includegraphics[width=0.45\linewidth]{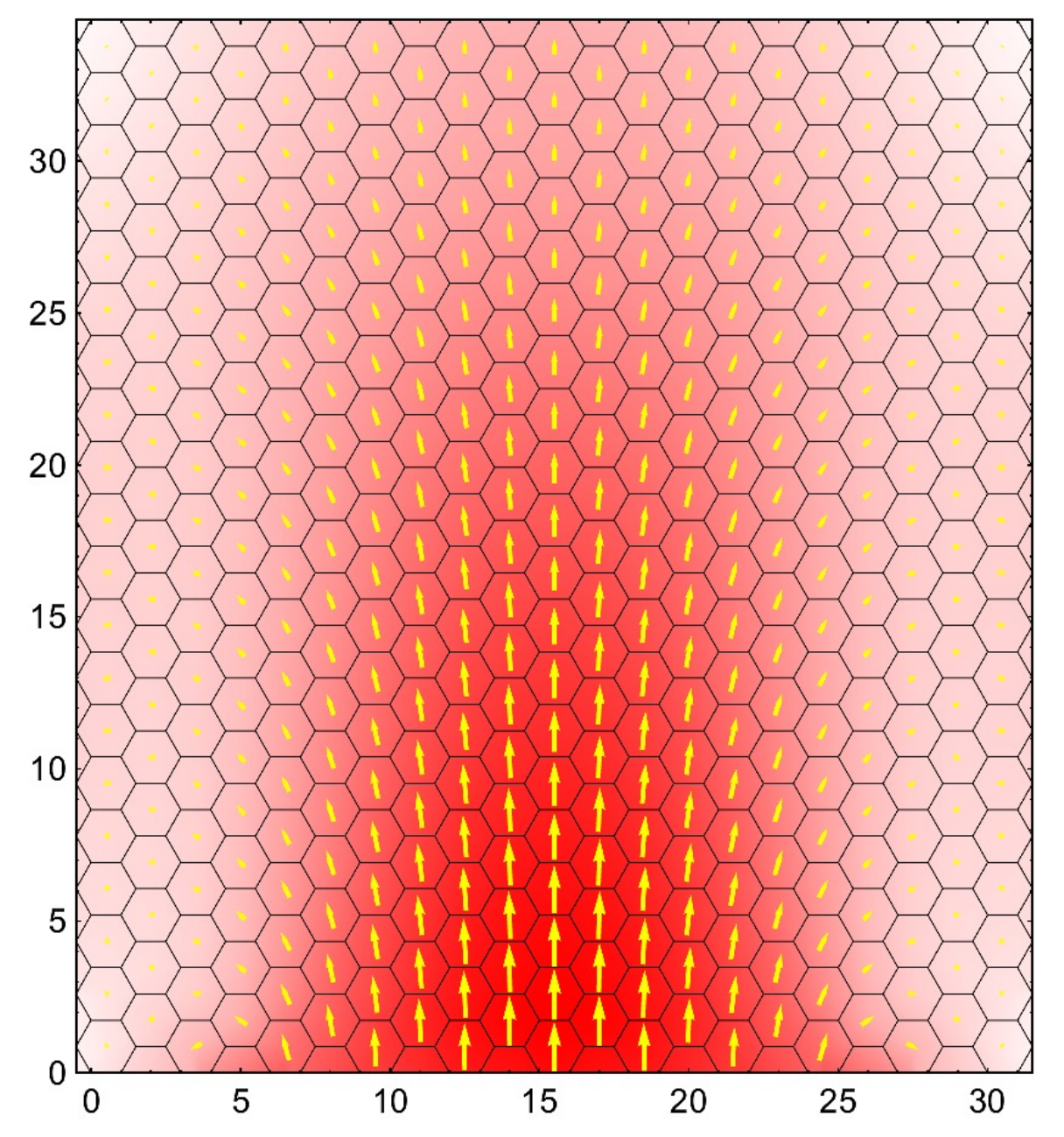}
  \includegraphics[width=0.45\linewidth]{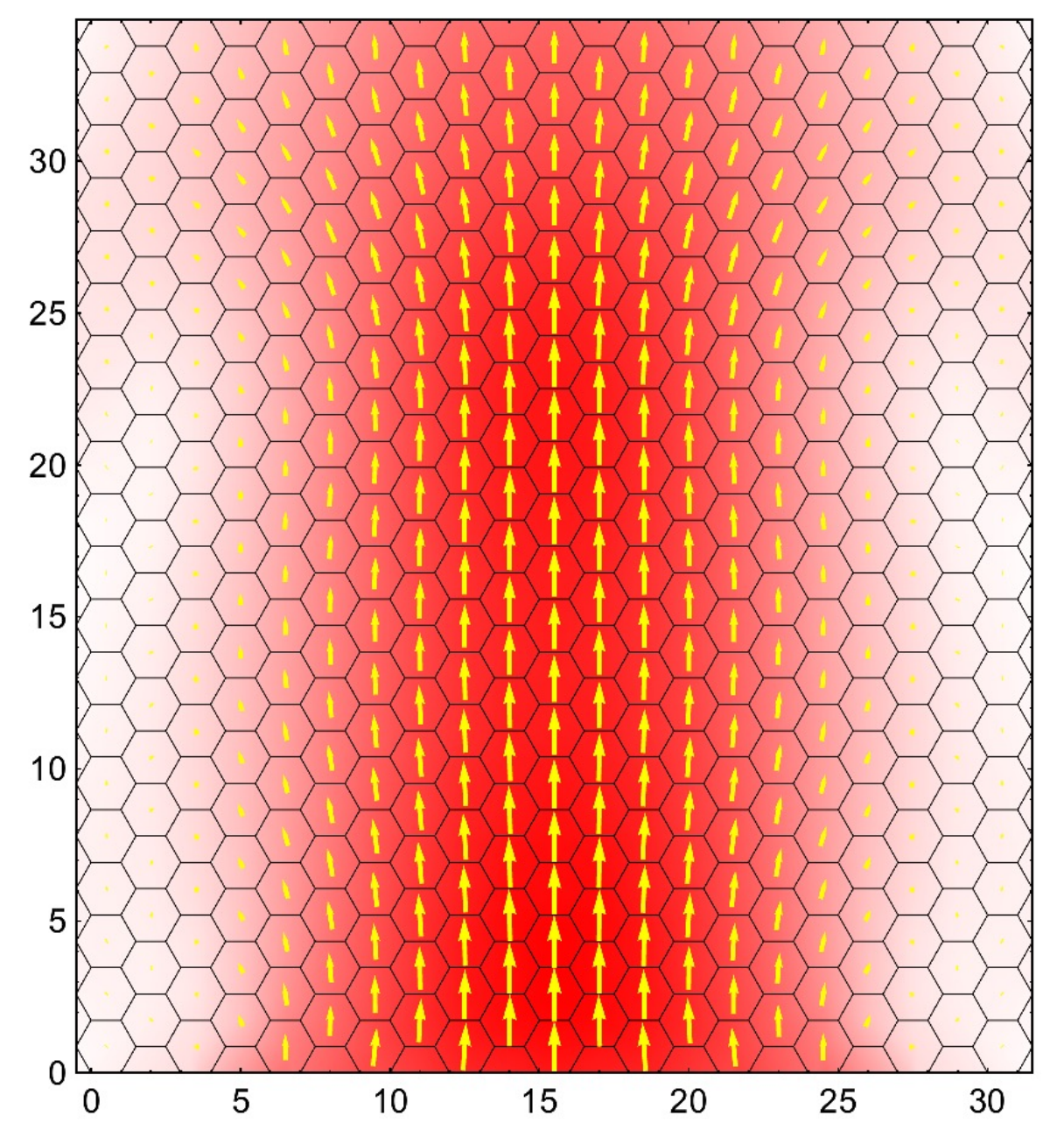}\\
  \includegraphics[width=0.45\linewidth]{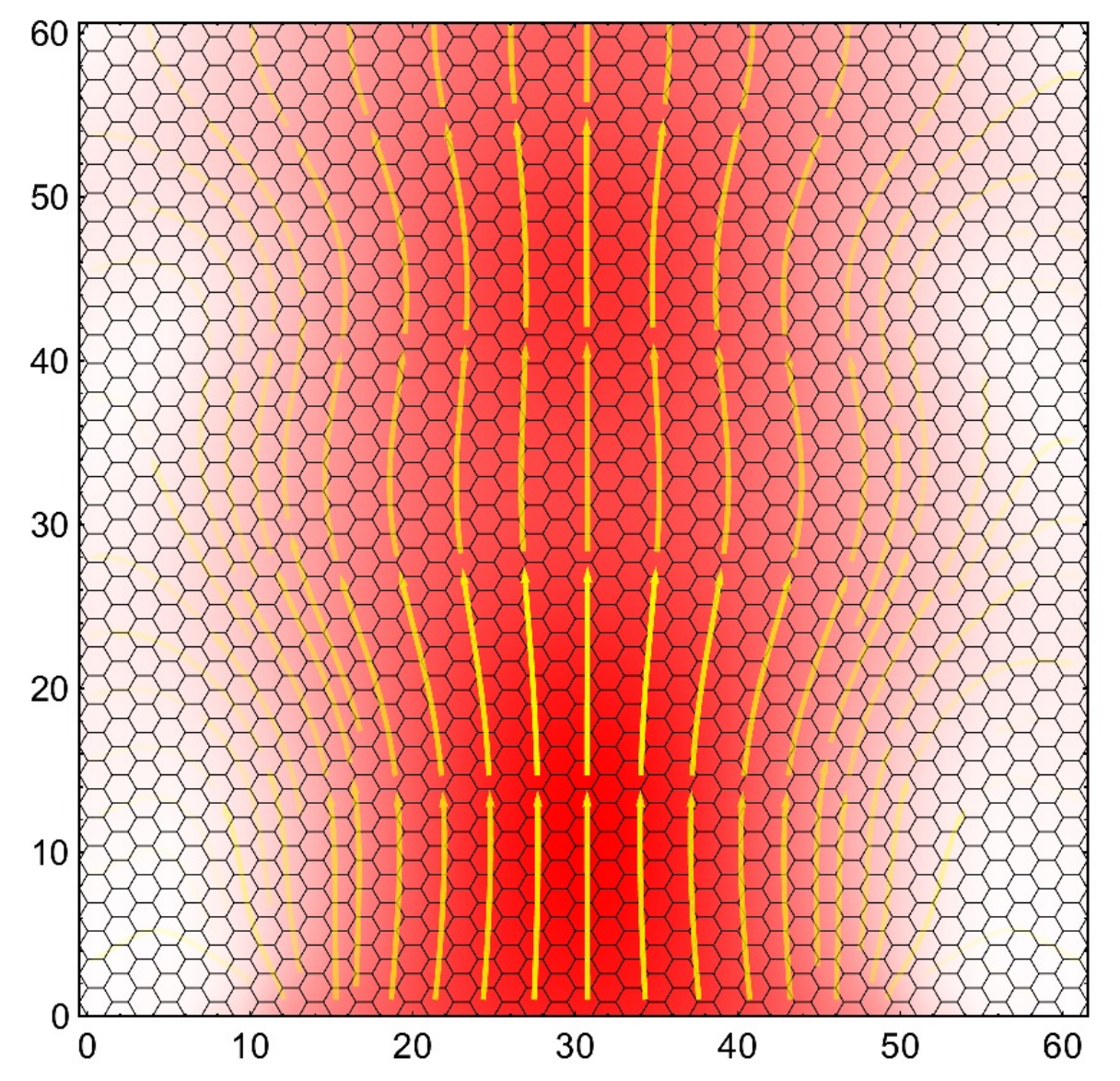}
  \includegraphics[width=0.45\linewidth]{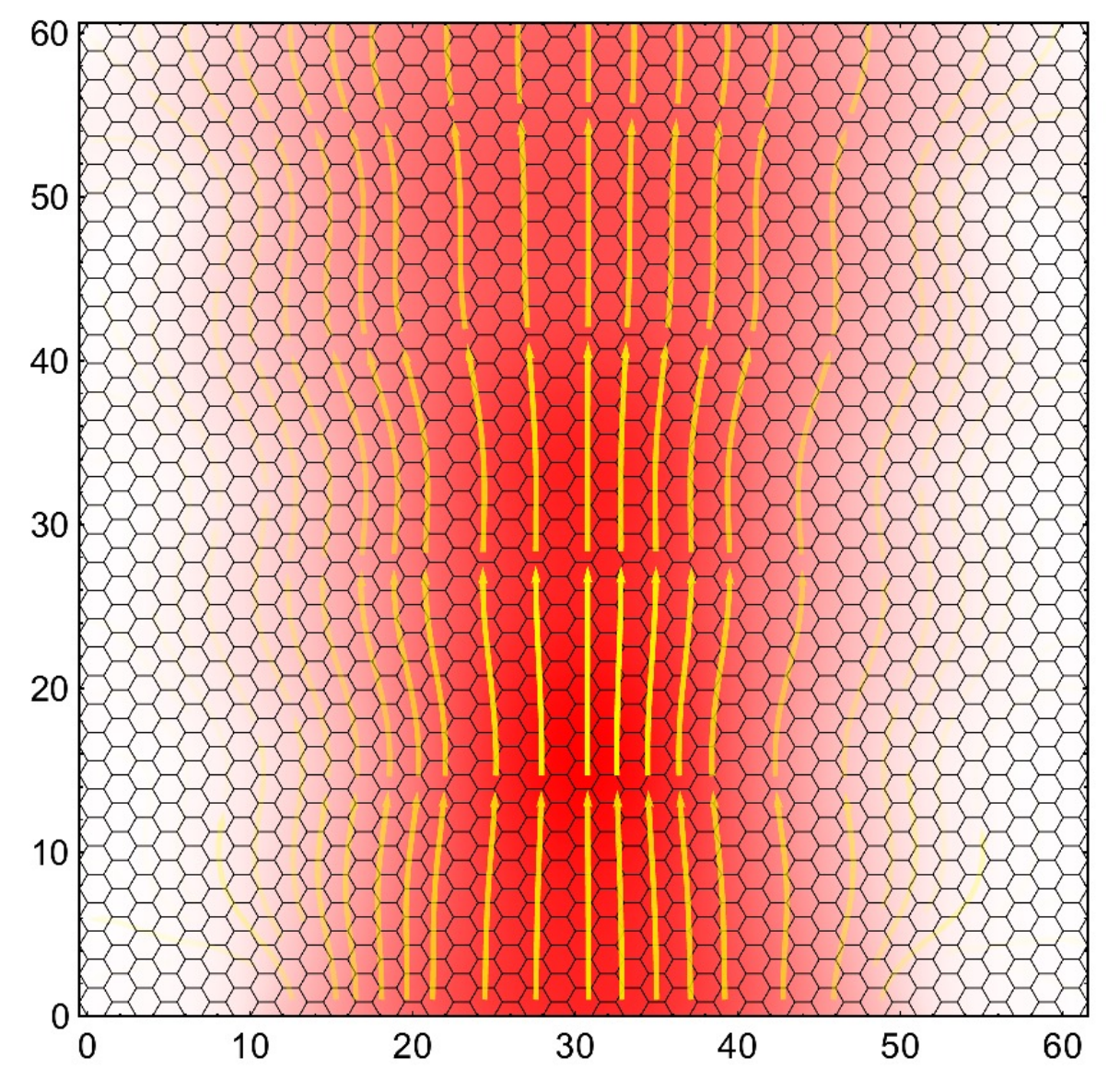}
  \caption{Injection of a plane wave at the contact at bottom edge. Top: current vector field pro hexagon (yellow vectors) and current density (red color shading) in a system of $21\times 20$ atoms. Bottom: current integral lines (yellow vectors) and current density (red color shading) in a system of $41\times 35$ atoms. Chosen energies: $E=0.2$ (left) and $E=0.3$ (right).
  \label{fig:Gauss-beam}}
\end{figure}

In order to obtain possibly narrow current flows we additionally give the injected ``beam'' a Gaussian form $|\psi^\text{in}(x)|^2 \sim \exp(-4\log(2)\,(x-x_0)^2/D^2)$ which is known for low spreading%
\footnote{Even better might be a Bessel form, which is non--diffractive, but we want to keep the beam's amplitude strictly positive. Gaussian and Bessel beams are used e.g. in laser techniques.}
and has half--width $D/2$
(cf. Fig. \ref{fig:Gauss-beam}).

\subsection{Pseudo-magnetic field} \label{sec:pseudo-B}

The action of geometry is fundamentally different from the action of a magnetic field when applied to particles and antiparticles.
Here, the electronic excitations above the Fermi level ($E>0$) behave as Dirac particles while the holes ($E<0$) behave as Dirac antiparticles with opposite charge. Both should react identically to curvature and oppositely to the pseudo--magnetic field.

There is, however, one problem in graphene: the number of Dirac excitations is doubled due to the existence of two inequivalent Dirac cones in the dispersion relation on the hexagonal lattice.
In deformed graphene, each of these two kinds ``feels'' the opposite sign of the pseudo--magnetic potential $\pm \v A$, which globally reflects the time--reversal symmetry present in the system.

The way out of this symmetric catch, implemented in our calculations, is an asymmetric injection of electrons at both valleys\footnote{Setups acting as valley polarizers have been discussed in \cite{Rycerz-ValleyFilter, Valleytronics, ValleyFilter+strain, Gunlycke-Graphene-LineDefect-ValleyFilter, LineDefect-ValleyPolarization}.}.
The contact formula \eqref{phase-factors} enables a simple filtering of valleys when the contact is placed parallel to the armchair edge and the injection momentum $\v k$ is chosen in the zigzag direction.
In such a case, \eqref{phase-factors} reduces to $\chi_{\v n} = C_- \pm C_+$ for $\v n\in\lA, \lB$, respectively (with $\sigma = +1$). By choosing $\chi_{\v n\in\lA}=\chi_{\v n\in\lB}$ we ensure injection only at valley $\K^-$.
The current flows then through the whole lattice mainly through that one channel --- projections onto the corresponding subspaces lead to amplitude ratios  $\K^{-}:\K^{+} > 10:1$ at all studied energies.

We observed that the form of the current flowing via valley $\K^-$ is more compact while the one via valley $\K^+$ is clearly wider.
At energies $E > 0.3$, the Dirac--cones become anisotropic, slightly triangular (cf. Fig. \ref{fig:G-Spectrum-pert}, left).
The flattened part helps the current to flow in one direction while the triangle--edge disperses the current away from the main direction.
At lower energies, $E < 0.3$, the dispersion relation is almost round and both currents, via $\K^\pm$, flow (without curvature)  almost identically.

\begin{figure}[ht]
  \includegraphics[width=0.7\linewidth]{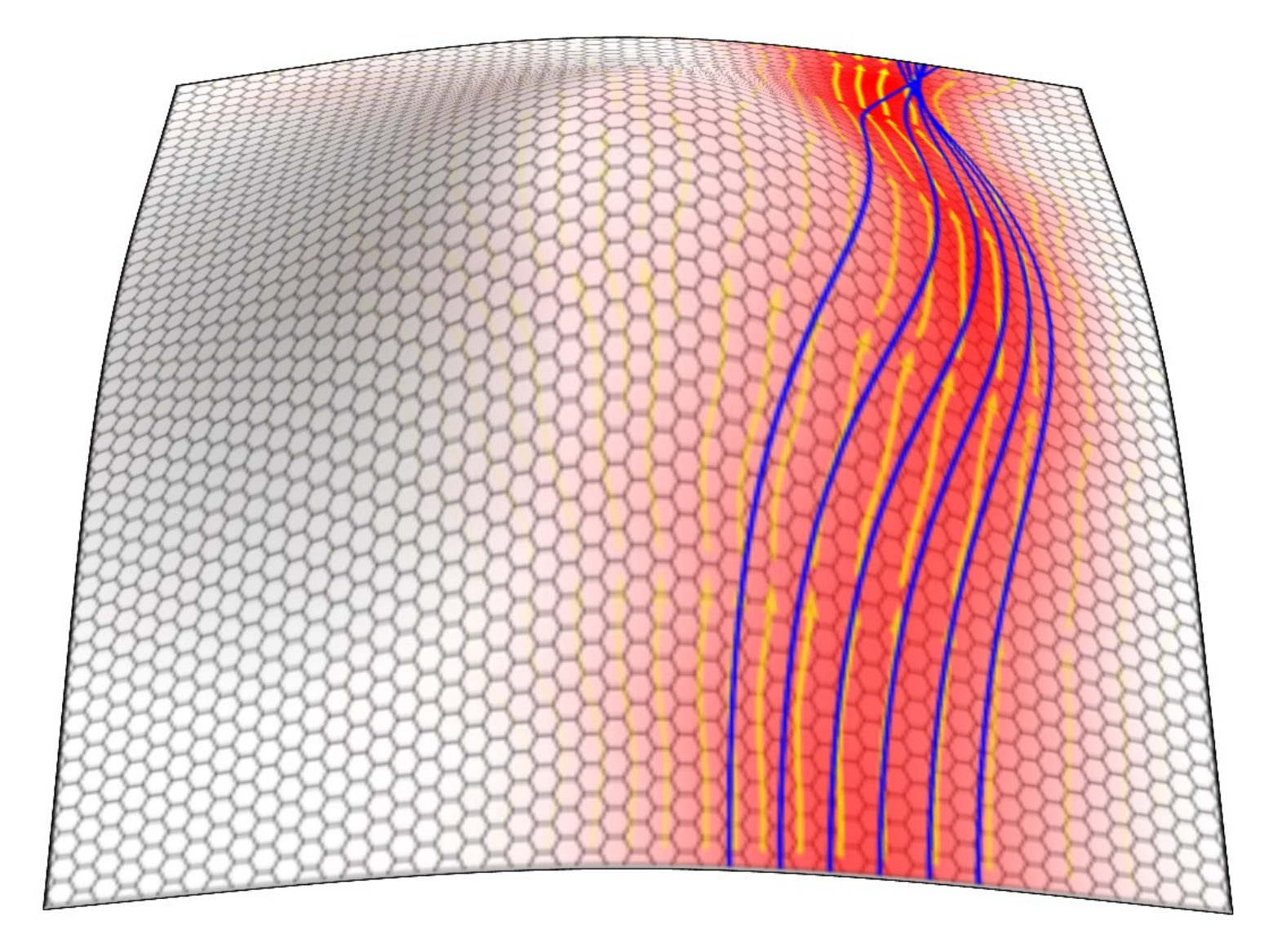}
  \caption{Electric current flow in curved graphene. Current stream\-lines (yellow arrows) and current density (red color shading)  calculated by the NEGF method compared to classical geodesics (blue solid lines) for massless charged particles moving in a magnetic field on the continuous curved surface.
  \label{fig:J+Geo+3D}}
\end{figure}

\begin{figure*}[t!]
  \subfloat[{$r_0=150\,d_0, h_0=0.50\,r_0$}]{\includegraphics[width=0.33\linewidth]{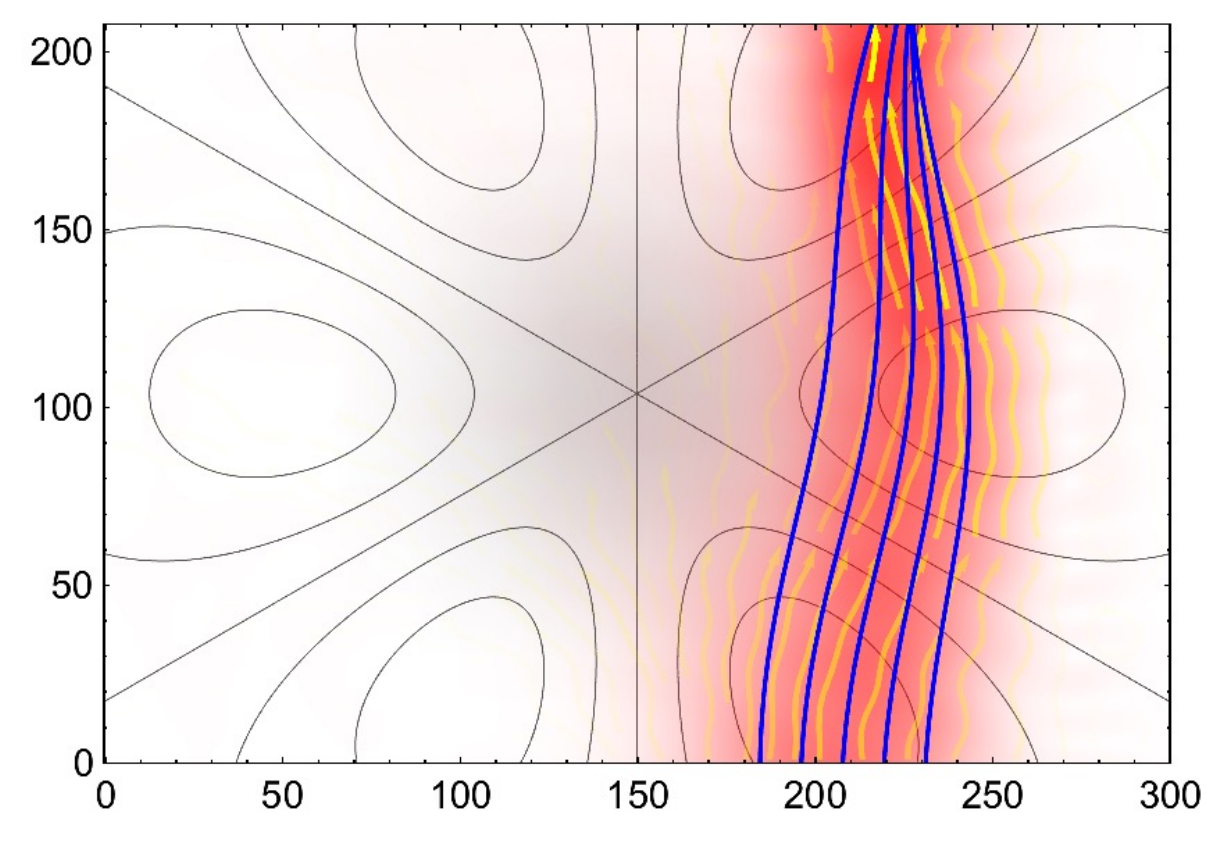}}
  \subfloat[{$r_0=150\,d_0, h_0=0.75\,r_0$}]{\includegraphics[width=0.33\linewidth]{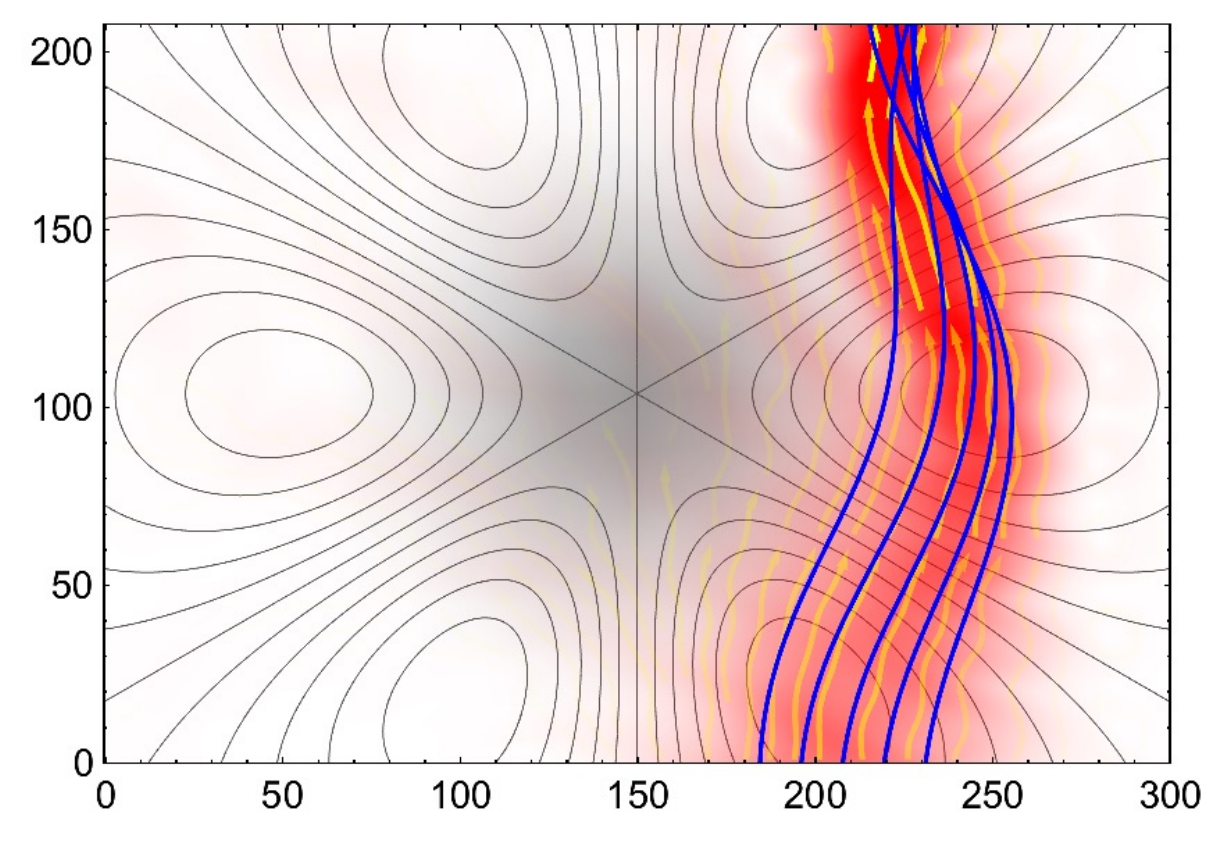}}
  \subfloat[{$r_0=150\,d_0, h_0=1.00\,r_0$}]{\includegraphics[width=0.33\linewidth]{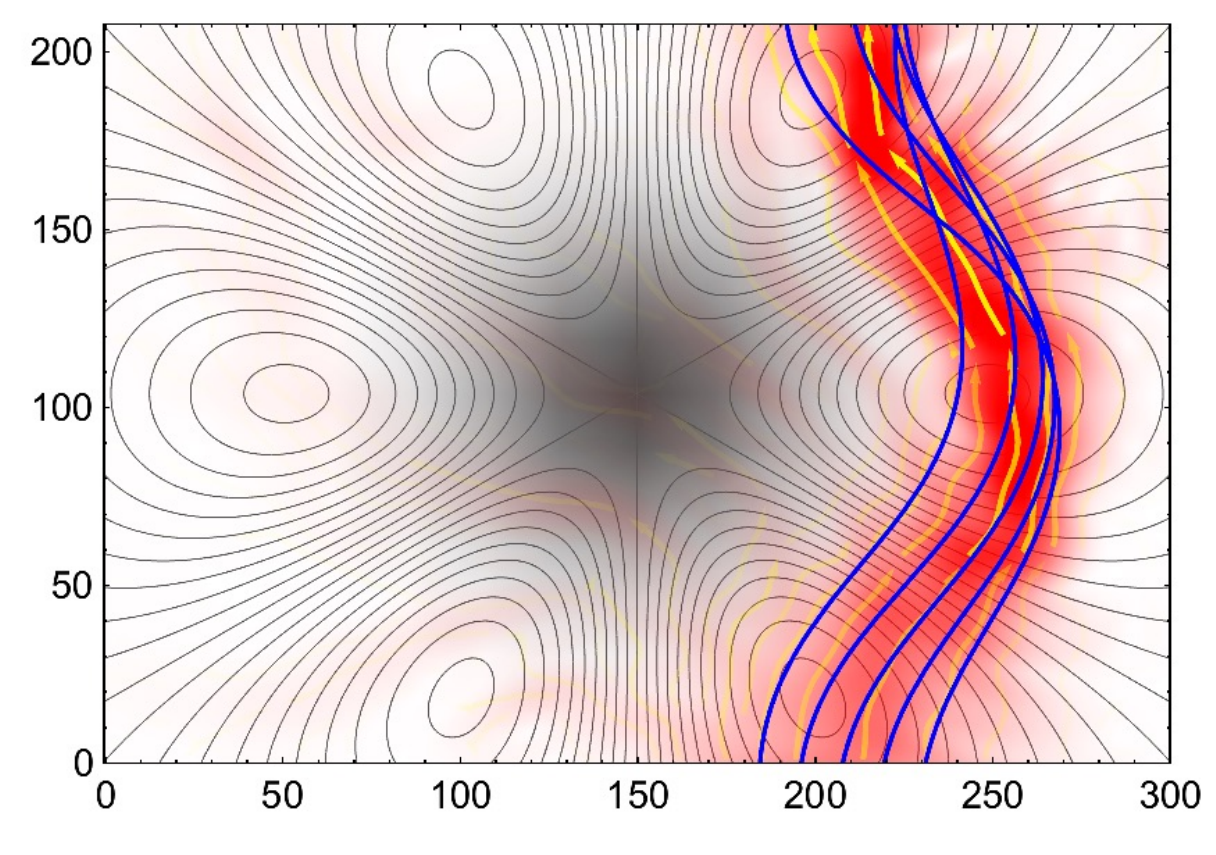}}
  \caption{
  Electric current flowing at energy $E=0.2\,t_0$ and optimal width (minimal spreading) through deformed graphene ribbons with varying height $h_0$ of the bump.
  The background (red) color represents current density with thin (yellow) current lines on top of it.
  Blue solid lines show classical geodesics of massless charged particles moving in the presence of a pseudo--magnetic field 
  (isolevels vary by $50\ \text{T}$) 
  on the curved surface.
  In (c) $\t B_\text{max} = 516\ \text{T}$.
  \label{fig:cur+geod_E=0.2}}
\end{figure*}

A typical current flow around a bump, from a contact at the bottom edge towards an opposite boundary is visualized in Fig. \ref{fig:J+Geo+3D}.
The maximal value of the pseudo--magnetic field, with spatial distribution as shown in Fig. \ref{fig:Struct+Ricci+A+B} (bottom), is given by $B_\text{max} = 0.377\, B_0\, \beta\, h_0^2\, r_0^{-3}$ with $B_0 = h/(e\, d_0^2) = 205\,100\,\text{T}$. 
In the examples discussed below, this value varies in the range $100 - 1000 \,\text{T}$ which is consistent with $300\,\text{T}$ estimated from the Landau levels for nanobubbles of similar size ($10\,\text{nm} \approx 70\,d_0$) \cite{Guinea+CastroNeto-StrainInduced300T}.

In the next Section, we discuss the influence of geometry on the current flows for various choices of parameters.

\section{Electric currents around a bump}

\subsection{Bending of current lines}

In the following examples we demonstrate the influence of different geometric factors on the shape of the current flow and its agreement with classically predicted geodesic lines.
In Fig. \ref{fig:cur+geod_E=0.2} the current is always injected at energy $E=0.2$ while the amplitude $h_0$ of the central deformation varies between $0.5$ and $1.0$ times the size of the deformation given by $r_0=150$.
The shape of the contact is chosen to be optimal according to the diffraction formula given above to keep the width of the current narrow along its whole path.

\begin{figure}[ht]
  \subfloat[{$E=0.2\,t_0, r_0=200\,d_0, h_0=r_0$}]{\includegraphics[width=0.7\linewidth]{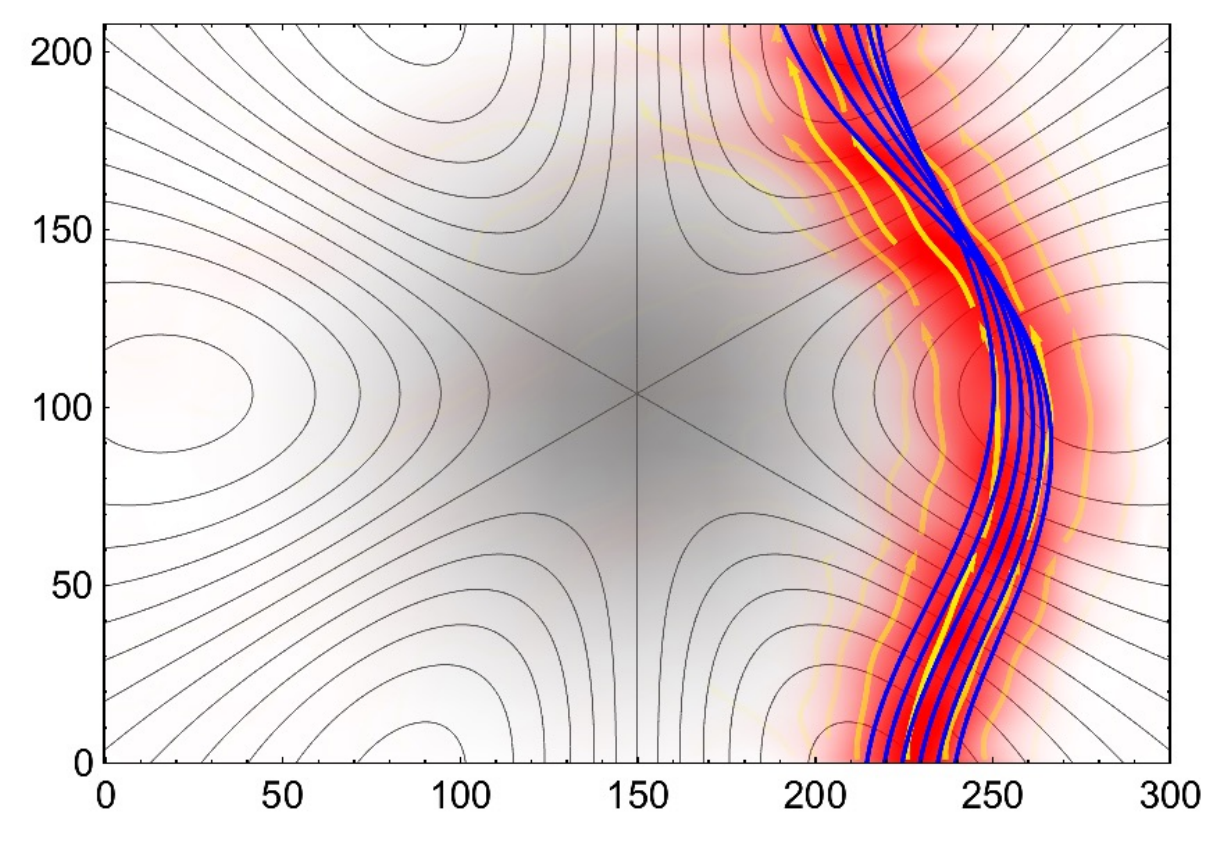}}\\ 
  \subfloat[{$E=0.3\,t_0, b=200\,d_0, h_0=1.25\,r_0$}]{\includegraphics[width=0.7\linewidth]{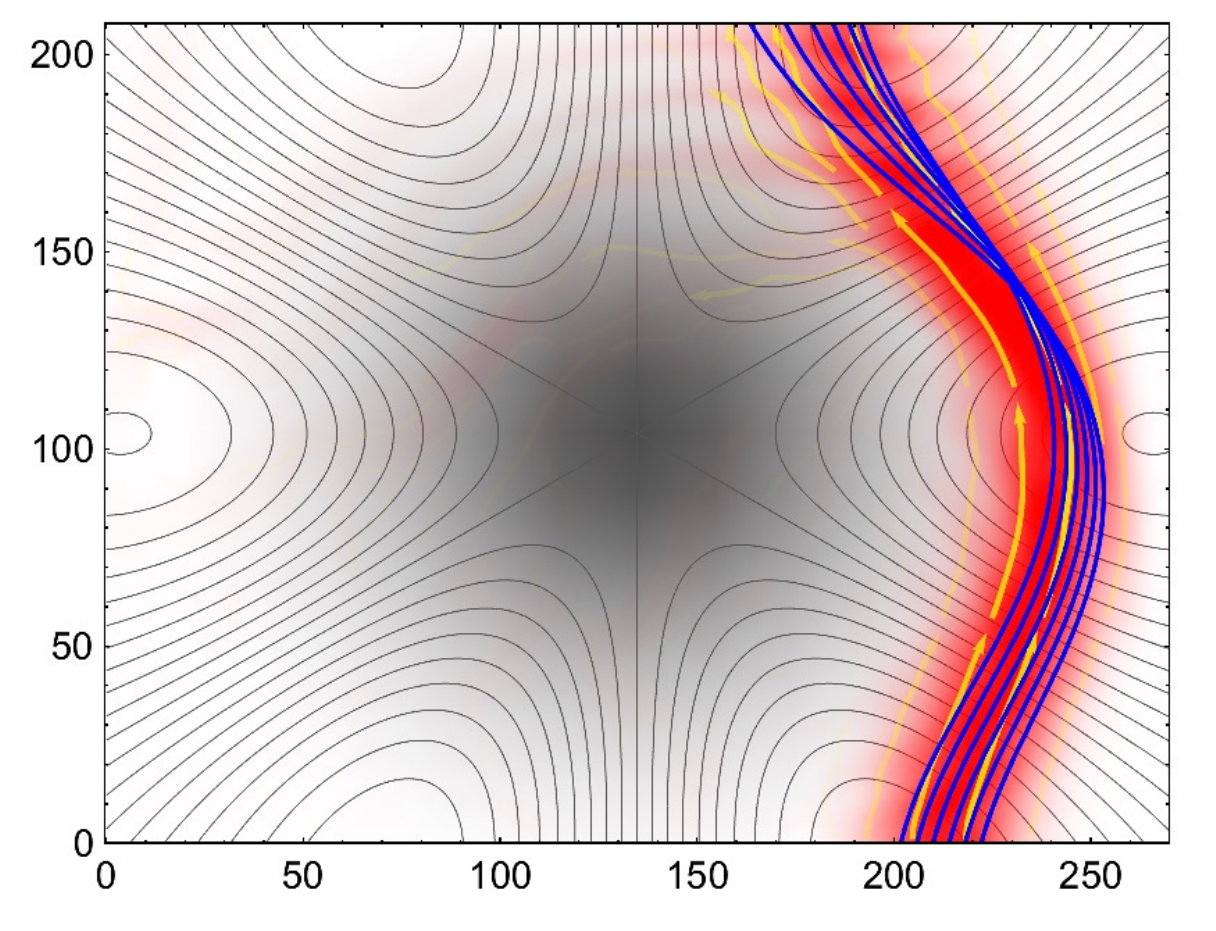}}\\ 
  \caption{Current flowing with narrower (by half) than predicted optimal width without spreading. Crossing of geodesics compensates for the expected diffraction. 
  (Further description as in Fig. \ref{fig:cur+geod_E=0.2}, isolevels by $60\ \text{T}$.)
  \label{fig:cur+geod_E=0.3_d=0.5}}
\end{figure}

However, the effect of crossing of geodesics helps to focus the current and leads to a narrower flow than what can be expected from the standard diffraction.
Fig. \ref{fig:cur+geod_E=0.3_d=0.5} shows two such examples at $E=0.2$ and $E=0.3$ where the contact width has been reduced by half and the current keeps its narrow width along the whole path while without the geodesic focusing a widening by factor 4 would be expected at the upper boundary.
The enhanced focusing takes place at the cost of slight disagreement appearing between the flow and the geodesics.
Its origin lays in the breakdown of the eikonal approximation relating waves to the current lines at caustics, i.e. where geodesics cross.

\begin{figure*}[ht]
  \subfloat[{$E=0.3\,t_0, r_0=100\,d_0, h_0=r_0$}]{\includegraphics[width=0.33\linewidth]{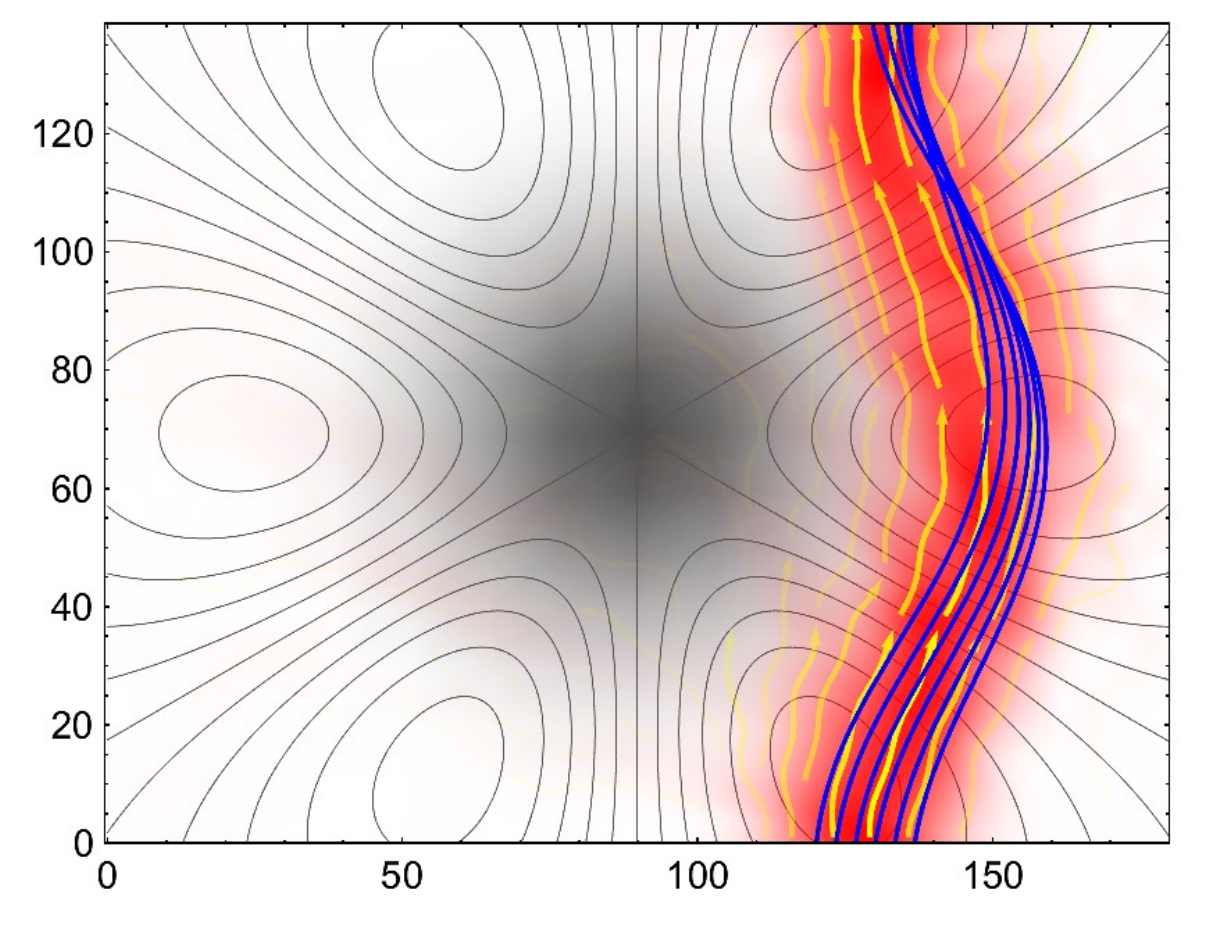}} 
  \subfloat[{$E=0.5\,t_0, r_0=100\,d_0, h_0=r_0$}]{\includegraphics[width=0.33\linewidth]{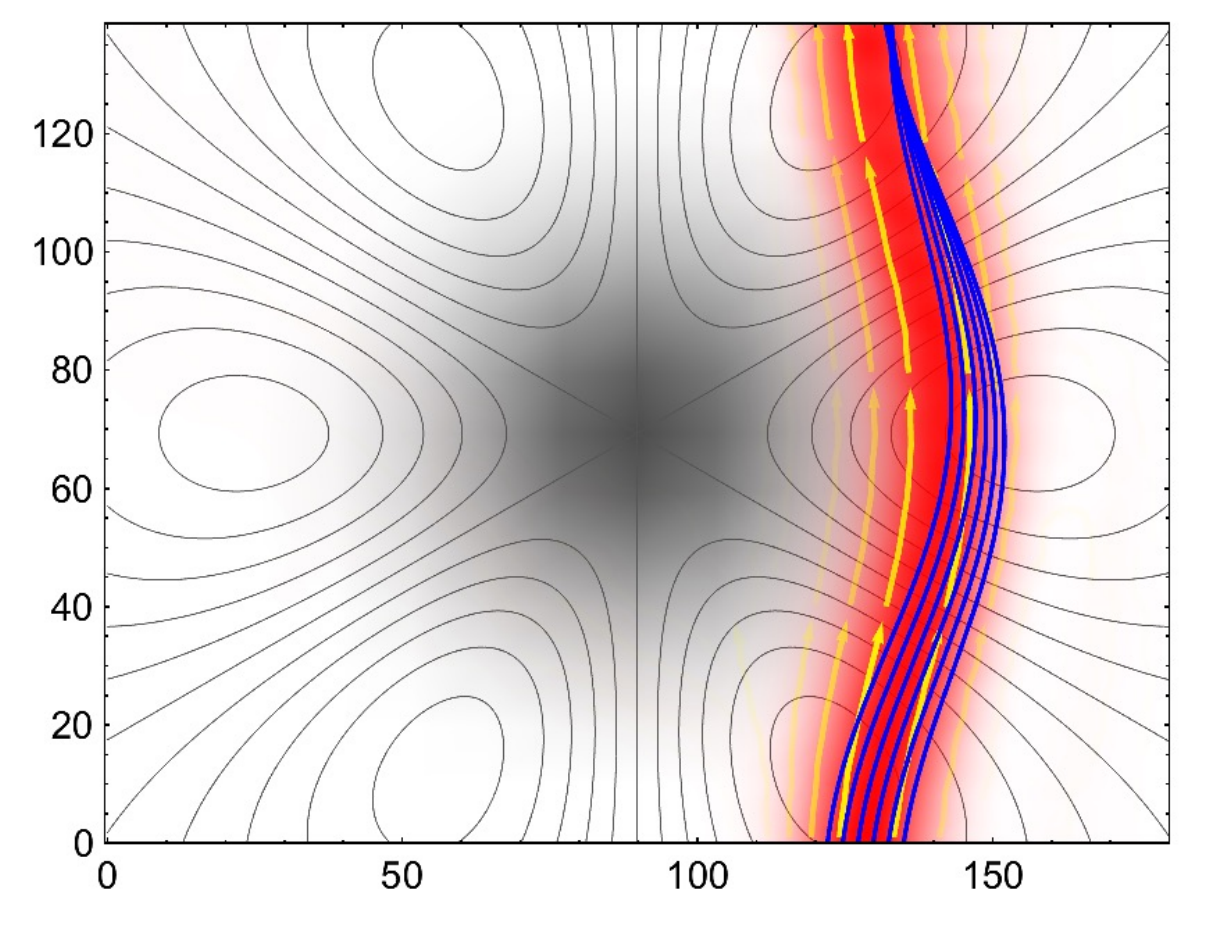}} 
  \subfloat[{$E=0.7\,t_0, r_0=100\,d_0, h_0=r_0$}]{\includegraphics[width=0.33\linewidth]{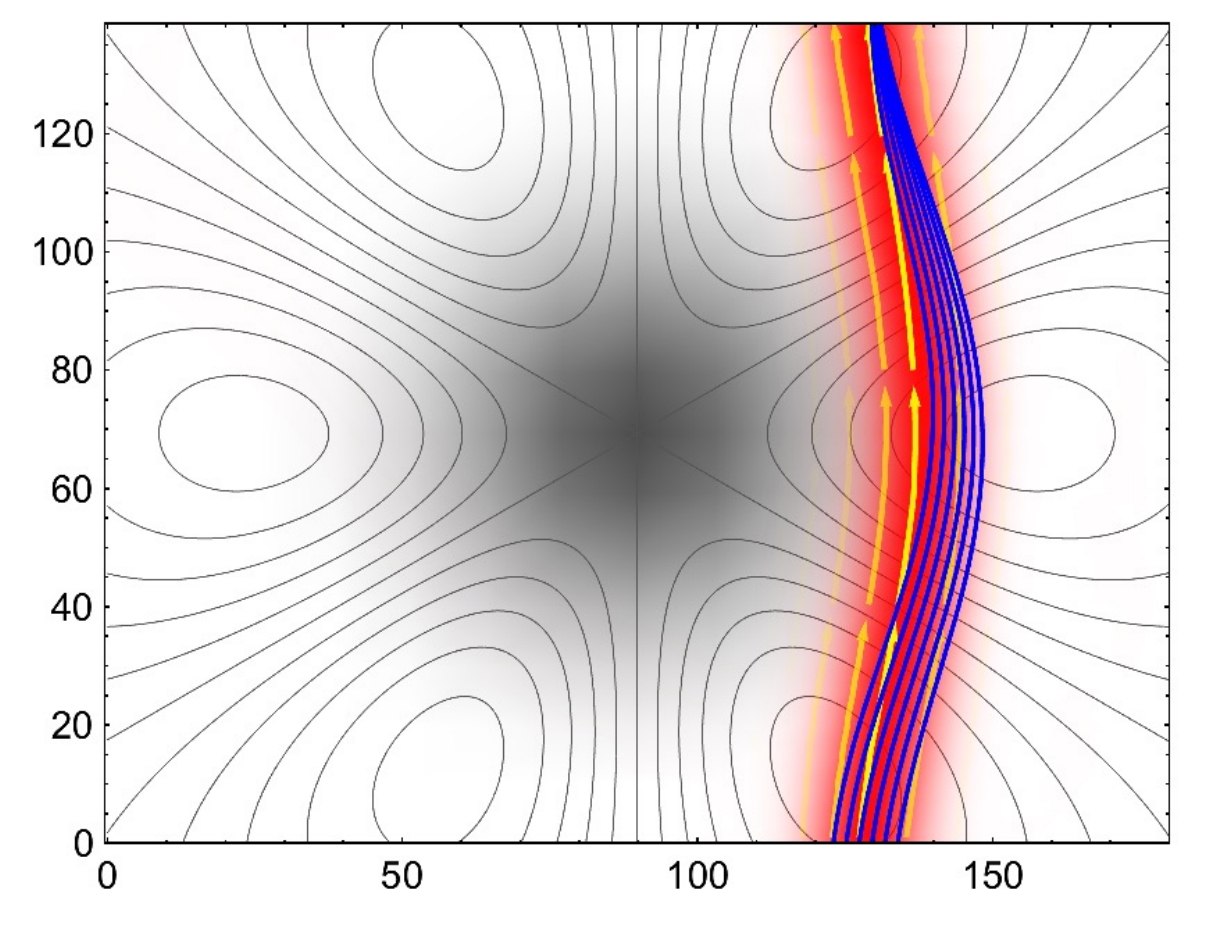}} 
  \caption{
  Electric current flowing through bump--like deformed graphene ribbons. 
  Left to right: varying energies $E$. Deformation of Dirac cones at $E\gtrsim 0.3$ leads to slight deviation from geodesics.
  \label{fig:cur+geod_Es}}
\end{figure*}

The last example, presented in Fig. \ref{fig:cur+geod_Es}, compares flows of current injected at different energies $E=0.3, 0.5$ and $0.7$ in the same geometry.  In each case, optimal contact width has been chosen in an energy dependent way. 
With increasing energy the influence of the pseudo--magnetic field becomes less relevant and the flow becomes visibly straighter.
The anisotropy of the Dirac cones at $E\geq 0.3$ leads for each valley to propagation in three dominant directions \cite{GreensF_Graphene}. Only the one pointing straight up contributes to the flow from the source located at the bottom edge (the other two point downwards). This explains the mismatch with geodesics for which an isotropic propagation is assumed.

\subsection{Behavior of particles and antiparticles}

The transformation  $\psi_\lB \ra -\psi_\lB$, which changes the sign of the wavefunction $\psi$ on the whole sublattice $\lB$ (while preserving the values of $\psi_\lA$), is a symmetry of the hopping Hamiltonian \eqref{eq:Ham-hopping}. Under its action, $\op{H}$ changes its overall sign, independently of whether the hopping terms are all equal or perturbed.
This enables us to generate solutions with negative energies $E<0$ from solutions with positive $E>0$ in a simple way.

\begin{figure}[ht]
  \includegraphics[width=0.49\linewidth]{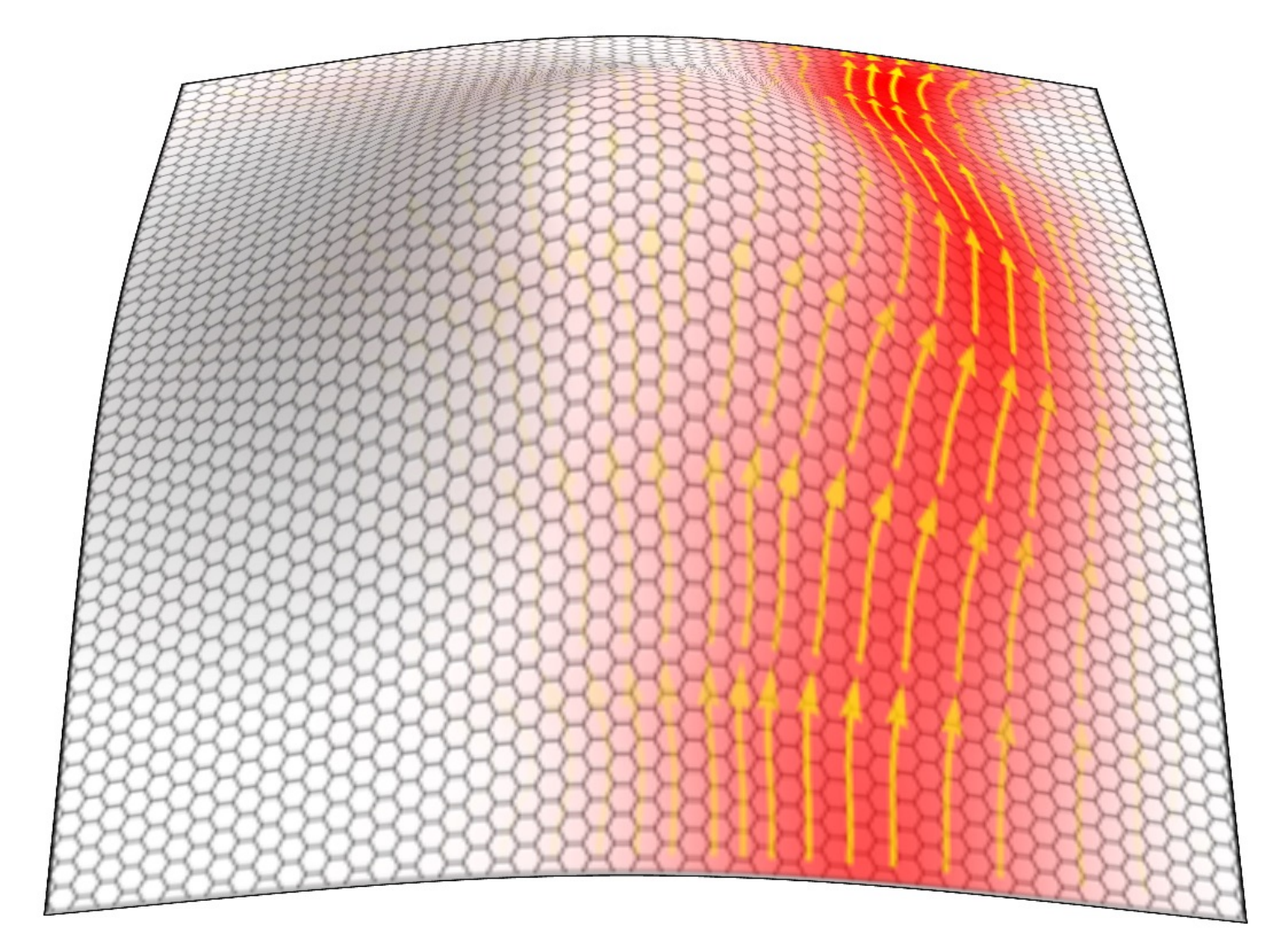}
  \includegraphics[width=0.49\linewidth]{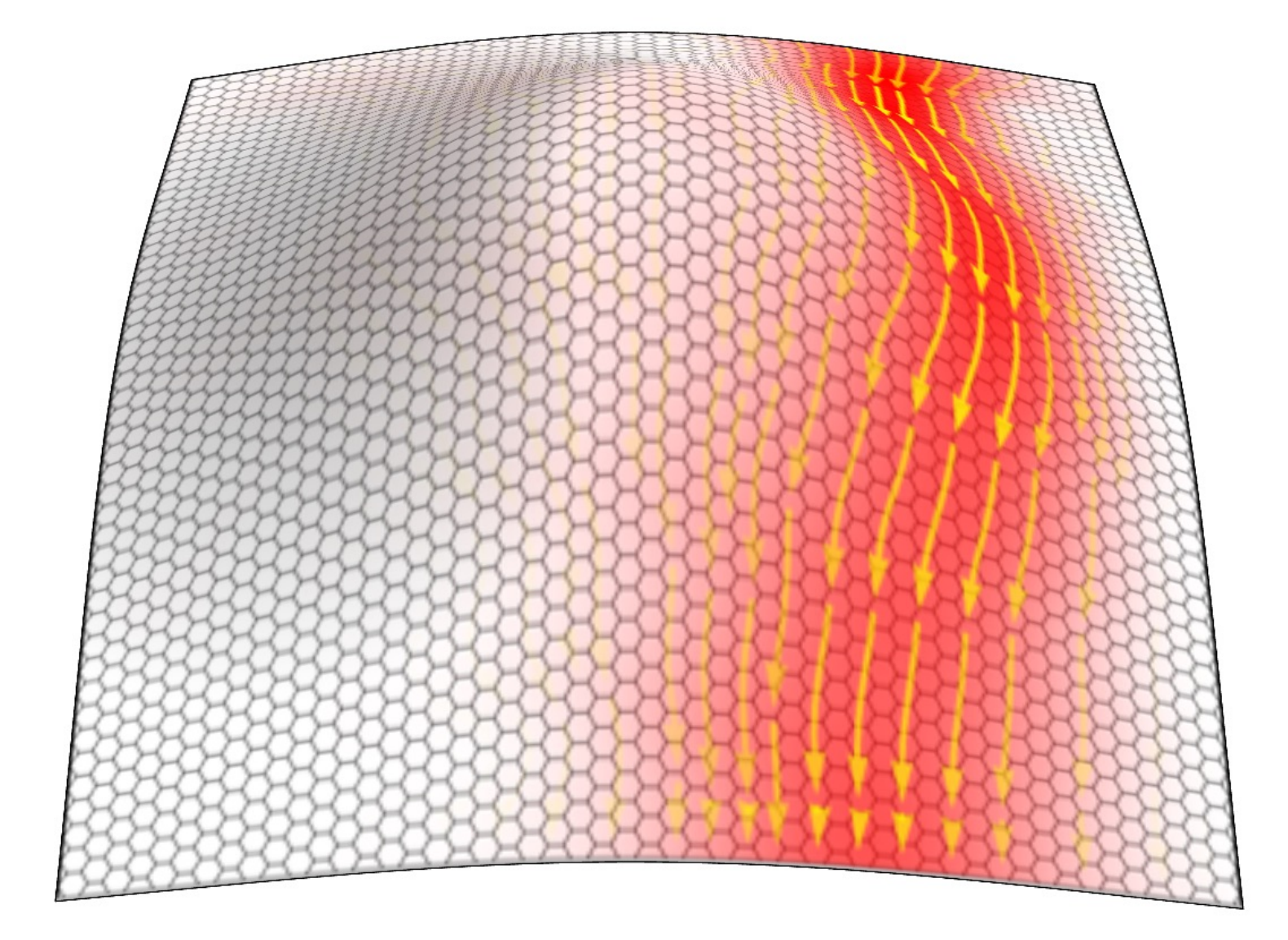}\\
  \caption{Currents of particles with $E=E_0>0$ (left) and antiparticles with $E=-E_0<0$ (right) follow identical paths as they react to the pseudo--magnetic fields $\t B$ of opposite signs and equal curvature.
  Since for antiparticles the current $\v I$ is anti--parallel to the pseudo--momentum $\v k$ the flow can be also interpreted as a flow of particles moving in the opposite direction.
  \label{fig:particles-antiparticles}}
\end{figure}

Performing numerical computations, we observed that the antiparticle (hole) current with $E=-E_0<0$ follows the same path as the corresponding particle current with $E=E_0>0$ and all other conditions remain unchanged. The only difference is that the current \eqref{eq:current-NEGF}, \eqref{current-lattice} changes its overall sign, i.e. flows in the opposite direction (cf. Fig. \ref{fig:particles-antiparticles}). This is a direct consequence of the transformation $\psi_\lB \ra -\psi_\lB$ or, in other words, of the time--reversal symmetry.

The time--reversal symmetry enforces also the transformation $\v k \ra - \v k$ (while keeping the same propagation direction) which exchanges the valleys $\K^-\ra\K^+$  and, in consequence, the sign of the pseudo--magnetic potential $\v A \ra -\v A$. Therefore,
the antiparticles (Fig. \ref{fig:particles-antiparticles}, right) ``feel'' the opposite pseudo--magnetic field $-\t B$ and
their current injected at the same contact and in the same direction follows the same path as that of particles (Fig. \ref{fig:particles-antiparticles}, left).

\begin{figure}[ht]
  \includegraphics[width=0.7\linewidth]{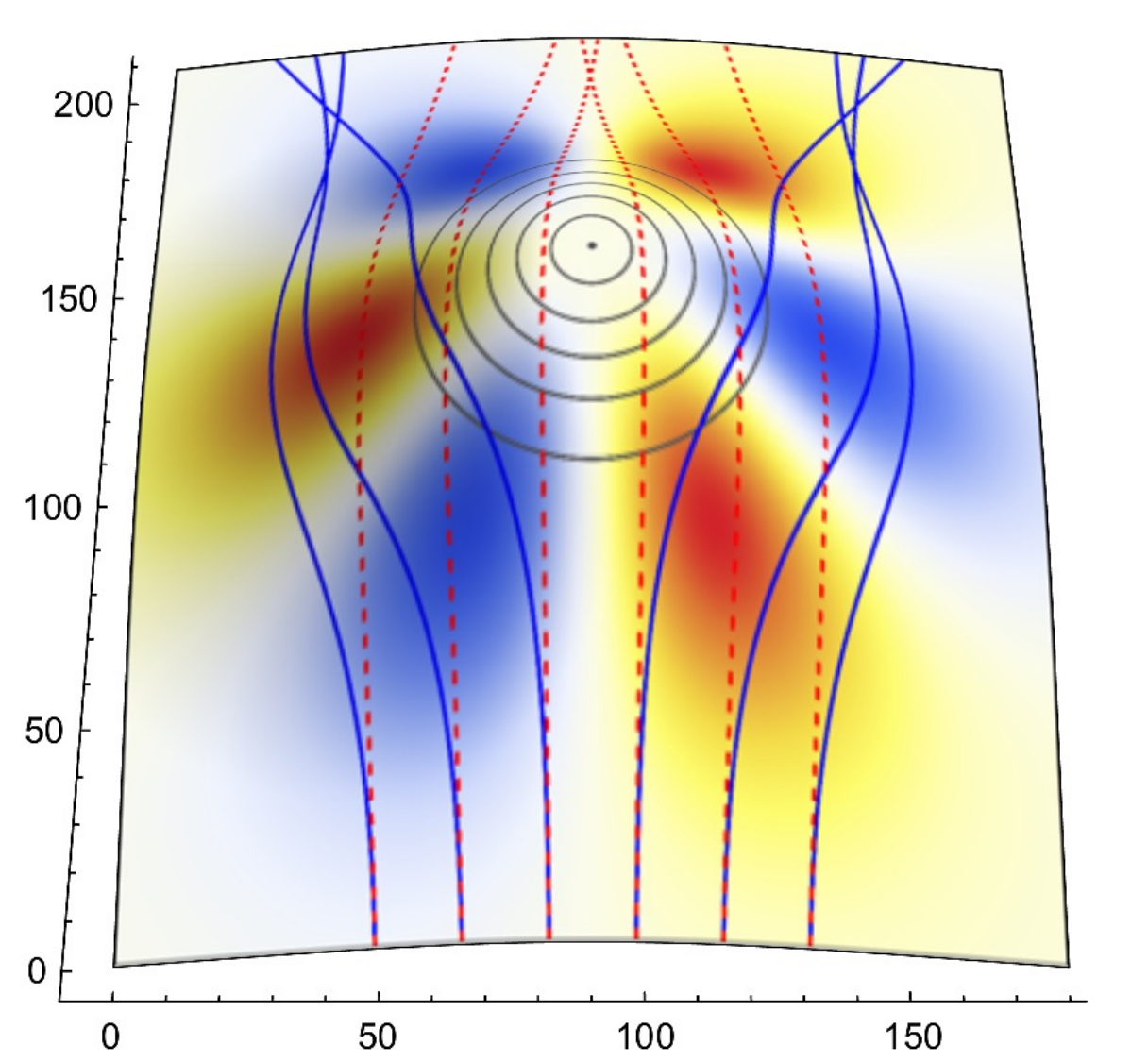}
  \caption{Geometric lensing: geodesic lines for curved geometry and pseudo--magnetic field (blue solid lines) compared to pure geometry without pseudo--magnetic field (red dotted lines, background shading as in Fig \ref{fig:Geo-GB0}).
  The pseudo--magnetic field shifts the focus away from behind the bump, creating two symmetric foci.
  \label{fig:Lens-idea}}
\end{figure}

\begin{figure*}[ht]
  \subfloat[$E=0.3\,t_0, r_0=70\,d_0, h_0=0.0\,r_0$]{\includegraphics[width=0.3\linewidth]{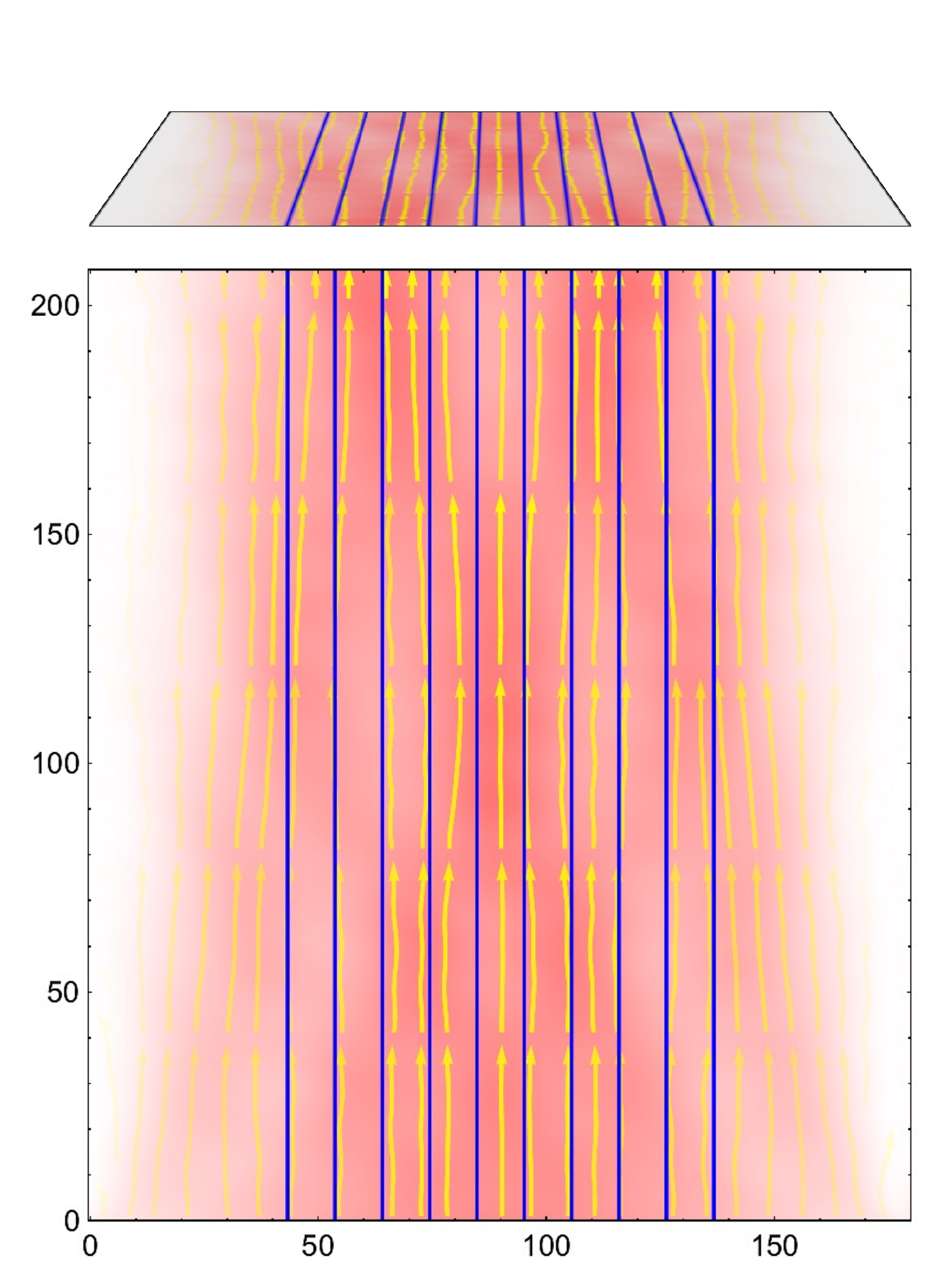}} 
  \subfloat[$E=0.3\,t_0, r_0=70\,d_0, h_0=0.5\,r_0$]{\includegraphics[width=0.3\linewidth]{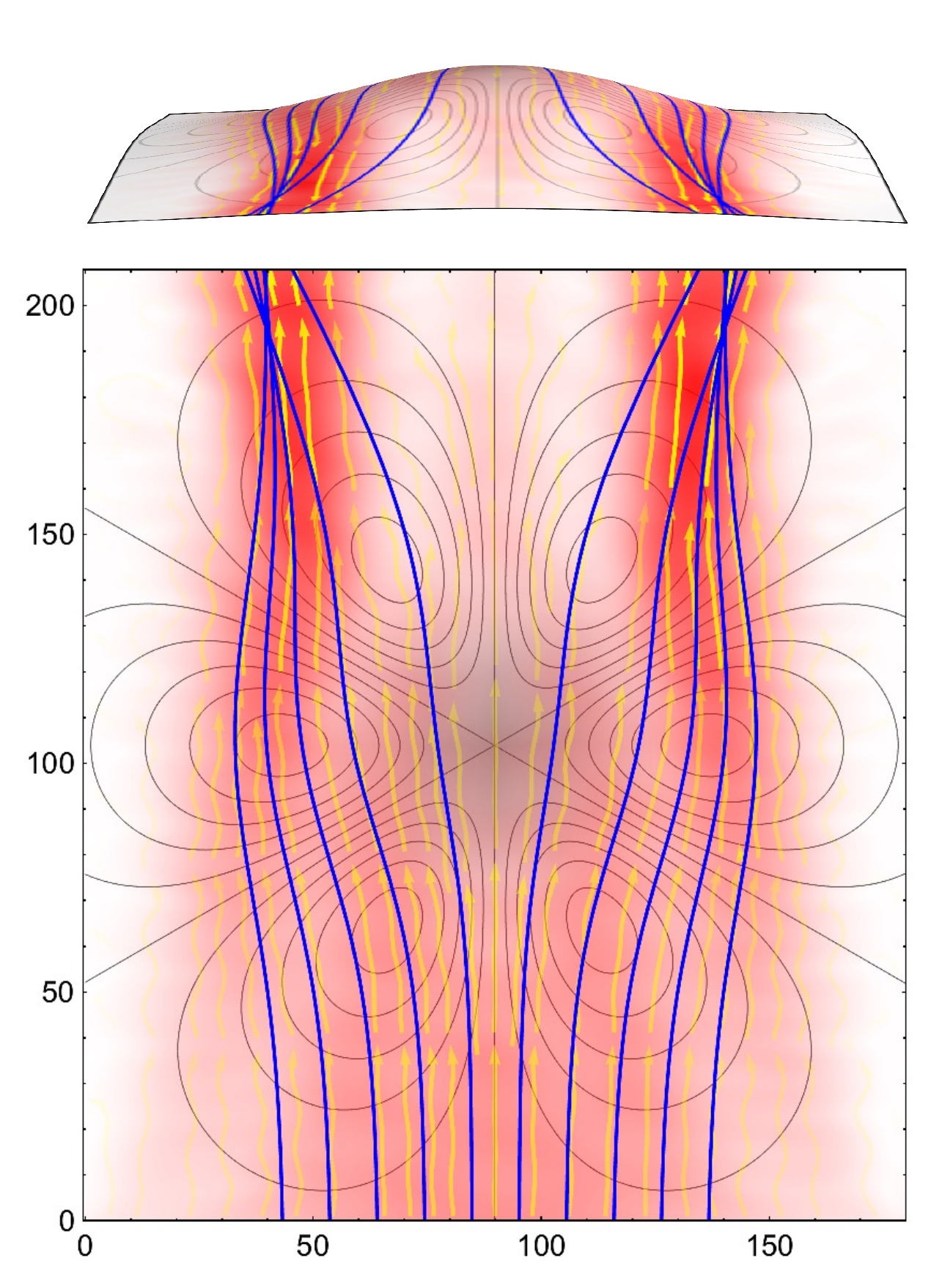}} 
  \subfloat[$E=0.3\,t_0, r_0=70\,d_0, h_0=0.7\,r_0$]{\includegraphics[width=0.3\linewidth]{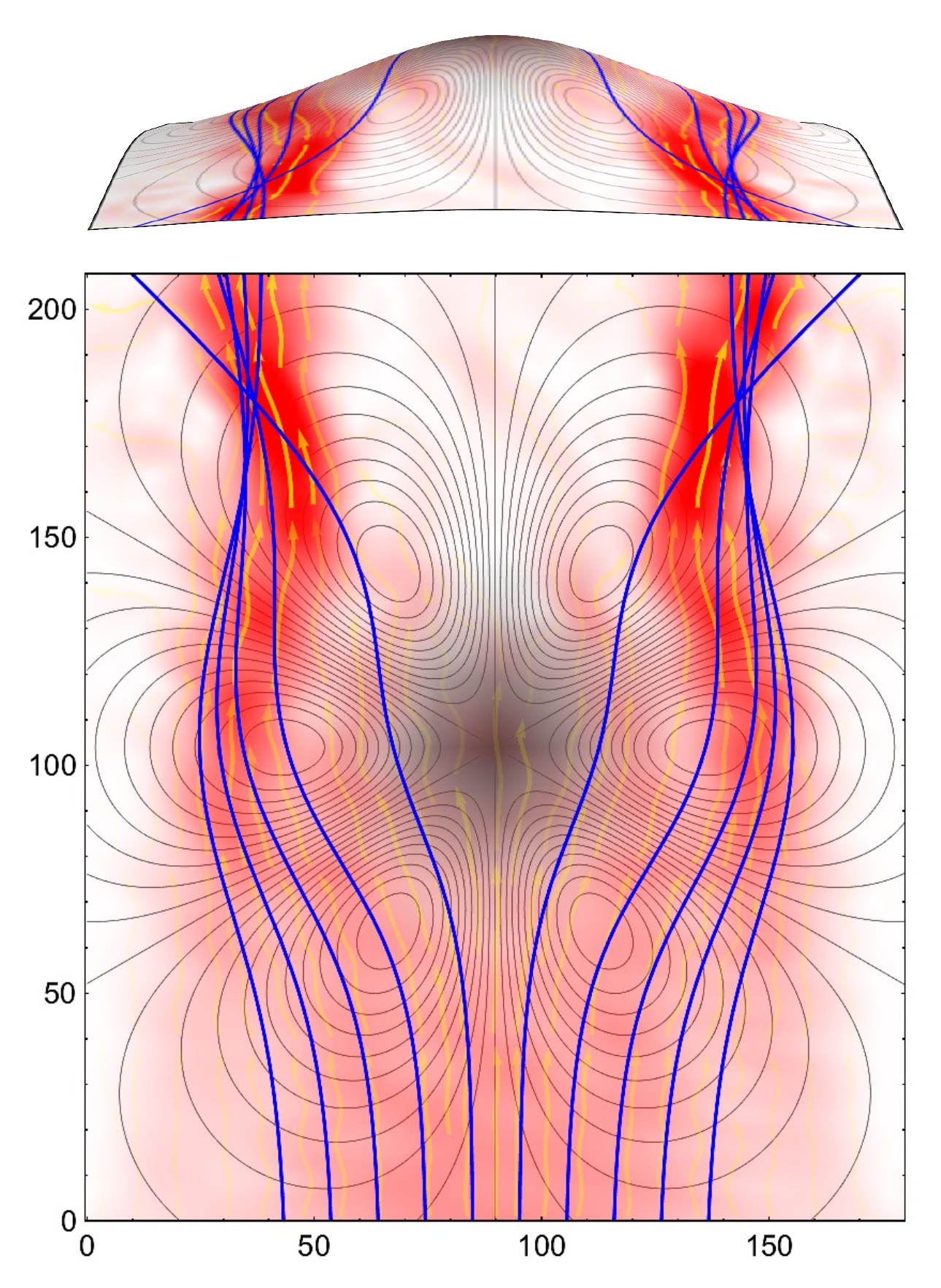}} 
  \caption{Electric current deflected and focused by a lensing geometry. Left to right: varying heights $h_0$ of the bump.  \label{fig:Lens-current}}
\end{figure*}

\subsection{Geometric lensing of currents}

One of the most intriguing applications of the presented ideas could be a purpose--built geometric lens that is able to deliberately focus or deflect electric current by an elastic deformation of the graphene surface.
The principle would be similar to the gradient--index lenses where the position dependent refractive index guides electromagnetic waves through the medium. Here, the position dependent metric and pseudo--magnetic field would guide the electron waves through graphene (cf. Fig.~\ref{fig:Lens-idea}) as has been discussed in \cite{GeodesicLenses, CurvedSurfaceLens} in the case of pure metric lenses. One such setup is presented in Fig.~\ref{fig:Lens-current}.
Note that the presence of the pseudo--magnetic field prevents the trajectories from crossing directly behind the bump. 
For the same reason closed geodesics encircling the bump and the corresponding quantum scattering effects \cite{GeodesicScatteringInTopInsulator, Guinea+Tagliacozzo-SpinConInTopInsulator} are not present here. 
Such a device might play a role as an ultra--sensitive pressure sensor built of graphene which would direct the electric current from a source contact (at the bottom edge) to one of several drain contacts (at the top edge) depending on the deformation of the surface. The differences between voltages measured at different contacts would provide information on the amplitude of the deformation.
We leave elaboration of this idea to a forthcoming publication.

\section{Discussion and outlook}

In this work we compared two fundamentally different approaches to the electronic transport in deformed graphene.
Firstly, using the NEGF method we computed the quantum currents \eqref{eq:current-NEGF} of electronic excitations directly from the hopping model \eqref{eq:Ham-hopping} with local modifications of the hopping parameters \eqref{eq:mod-hopping} due to strain. 
Secondly, we integrated classical trajectories \eqref{eq:geodesic} for relativistic charged massless particles moving in a curved 2-dimensional surface $z=h(x,y)$ in the presence of an emergent pseudo--magnetic field \eqref{eq:pseudomagneticF}.
The connection between the two approaches has been established via an effective 2-dimensional Dirac Hamiltonian in curved space \eqref{eq:Ham-Dirac-deformed} appearing in the long--wave limit from the special dispersion relation of graphene at low energies.
By applying geometrical optics approximation to focused plane wave beams we were able to switch from the wave to the particle picture in which the current lines are represented by the above mentioned classical trajectories.
We obtained very good numerical agreement between these pictures for a fairly wide set of parameters.
It has also been confirmed in time--dependent numerical simulations, similar to those in \cite{Covaci-WaveScatteringStrainedGraphene}, in which wave--packets of finite size have been sent through the deformed lattice. Their propagation has been consistent with the calculated classical trajectories \cite{Aurelien-private}.

The presented analogue model is based on approximations which are valid for the following hierarchy of scales
\begin{equation*}
  \setlength\fboxsep{0.2cm}
  \boxed{
  \begin{matrix} \text{lattice} \\ \text{constant} \end{matrix} \ll \text{wavelength} \, \ll \, 
  \begin{matrix} \text{deformation} \\ \text{scale} \end{matrix}
   \, < \, \begin{matrix} \text{system} \\ \text{dimensions} \end{matrix}
  }
\end{equation*}
accompanied by the narrow-flow condition
\begin{equation*}
  \setlength\fboxsep{0.2cm}
  \boxed{ 
  \text{optimal contact width} \; \sim \; \sqrt{\text{system size} \times \text{wavelength}} 
  }
\end{equation*}
which are satisfied for most real systems of interest%
\footnote{Wavelengths $30\, d_0 < \la < 3\cdot 10^4\, d_0$ for injection energies $10 - 1000\, \text{mV}$ ($0.003<E<0.3$, for $E \gtrsim 0.3$ significant deviation from conical form, for $E \lesssim 0.01$ room temperature noise), deformation scale of typical ripples and bumps $100 - 1000\, \text{nm} \sim 10^3\, d_0$, system size $\gtrsim 1\, \mu m \sim 10^4\, d_0$, contact widths $\gtrsim 100\, \text{nm}$.}.
Generally, the precision of the proposed approximations increases with the system size.

Clearly, the focus of the current paper has been put on the analogy between the distorted graphene and the Dirac equation in curved 2-dimensional space. This is why coherently injected plane--wave currents were of special interest.
However, beside the simulation of quantum fields in curved spaces,
the presented analogy has the potential of becoming an alternative, comfortable and efficient tool for calculating the electronic properties of newly designed graphene nanostructures.
It offers an enormous reduction of the complexity from irregular hopping Hamiltonians defined on large hexagonal lattices to the semiclassical geometric language for the description of curvature effects in a continuous surface.

For the proposed applications, the following extensions of the presented setup seem to be most natural.
Instead of injection at one contact, the current would rather flow between two or more defined contacts and take real boundaries into account.
An extended model of contacts can be included, allowing for injections of electrons at various grades of coherence.
A semiclassical description of the contact and boundary properties as well as of the interference effects for mixed currents flowing through both valleys, $K$ and $K'$, would be required.
Also the significance of corrections from the electron--electron interactions could be taken into account.

Summarizing, the main results of the current work include
the demonstration of the applicability of the continuous approximation in the prediction of current flow paths,
verification of the relative significance of the pseudo--magnetic and geometric effects
and the observation of the geometric lensing of currents in elastically deformed graphene.

\acknowledgements

T. S. acknowledges a postdoctoral fellowship from DGAPA--UNAM and financial support from CONACyT research grant 219993 and PAPIIT--DGAPA--UNAM research grants IG101113 and IN114014.
N.S. acknowledges financial support from the Deutsche Forschungsgemeinschaft (DFG).

\appendix
\section{Emergent fields from perturbed hopping parameters} \label{App:fields-from-hopping}

As already noted in Sec. \ref{sec:Dirac-from-lattice}, in a more general approach, the perturbations of the physical hopping parameters $\d t_{\v n,l}$ on the lattice can directly determine the (inverse) metric of the effective geometry 
\begin{equation}
  \t{g}^{ij}_{\v n} = \begin{pmatrix}
                 1 + \frac4{3}\left(\d t_{\v n,1} + \d t_{\v n,3} - \frac1{2}\d t_{\v n,2} \right), &
                 \frac{2}{\sqrt{3}} (\d t_{\v n,1} - \d t_{\v n,3}) \\[0.7em]
                 \frac{2}{\sqrt{3}} (\d t_{\v n,1} - \d t_{\v n,3}), & 1+2\d t_{\v n,2}
               \end{pmatrix}
\end{equation}
and of the pseudo--magnetic vector potential $\tK{}^{(s)}(\v x)=\K^{(s)} + \dK^{(s)}(\v x)$ where \cite{Katsnelson-GrapheneBook}
\begin{equation}
\begin{split}
   \dK^{(s)}_{\v n} &= (-1)^{s} \frac2{3}\, \underline{\bm{O}}_R \suma \da\ \d t_{\v n,l} \\
   &= (-1)^{s} \begin{pmatrix}
        \frac1{\sqrt{3}} (\d t_{\v n,1} - \d t_{\v n,3}) \\[0.7em]
        \frac1{3}(2\d t_{\v n,2} - \d t_{\v n,1} - \d t_{\v n,3})
     \end{pmatrix}
\end{split}
\end{equation}
($\underline{\bm{O}}_R$ is an operator of clockwise rotation by $\pi/2$ in the $x$-$y$ plane).
The effective strain is then given by
\begin{equation}
   \teps_{\v n} = \beta \left[ -\frac4{3} \suma \d t_{\v n,l} (\da \otimes \da) + \frac1{3} \left( \suma \d t_{\v n,l}  \right) \1 \right] = \beta \leps_{\v n}.
\end{equation}

\section{The discrete Dirac current} \label{App:discrete-Dirac-current}

The Dirac current in curved 2+1-dimensional space
\begin{equation}
  j^\mu = e^\mu_{\ \alpha}\, \psi\s \gamma^0 \gamma^a \psi \qquad (\mu,\alpha = 0,1,2)
\end{equation}
($\gamma^a$ are Dirac matrices in flat space) satisfies the covariant conservation law
\begin{equation}
  \nabla_\mu j^\mu = \frac{1}{\sqrt{g}} \p_\mu \left(\sqrt{g} j^\mu \right) = 0.
\end{equation}
Since the 2+1--metric is static and has the structure
\begin{equation}
  g_{\mu\nu} = \begin{pmatrix}
                  1 & 0  \\
                  0 & g_{ij}
               \end{pmatrix}
\end{equation}
the above current conservation law can be re-stated as
\begin{equation}
  \d_t \rho + \sqrt{g}^{-1}\p_k (\sqrt{g} j^k) = 0
\end{equation}
with $\rho = j^0$.
For stationary currents ($\p_t \rho = 0$) this conservation law reduces to $\p_k I^k = 0$ for $I^k = \sqrt{g}\, j^k$
or $I^k = \sqrt{g}\, e^k_{\ a} j^a$ where $j^a = \psi\s \sigma^a \psi$ with the choice $\gamma^0 \gamma^i = \sigma^i$ (and $\gamma^0 = \sigma^3$).

The last object can be conveniently discretized on the lattice where the components of $\psi=(\psi_\lA,\psi_\lB)$ are defined on the two sub-lattices $\lA$ and $\lB$ and gain additional phase--factors $\exp(-i \v K^{(s)} \v r)$ when expanded around a given Dirac $\v K^{(s)}$--point.
This gives a current defined on the lattice links (between sites $\v n$ and $\v m$)
\begin{equation} \label{current-lattice}
  j^a \ra J_{\v n, \v m} = i (\psi_{\v n} \psi\sc_{\v m} - \psi\sc_{\v m} \psi_{\v n}).
\end{equation}
Discretization of the metric determinant and frame vectors gives additionally $\sqrt{g}\, e^k_{\ a} \ra t_{\v n,\v m}$.
Both together give the discrete conserved current
\begin{equation}
  I^k \ra I_{\v n,\v m} = t_{\v n,\v m} J_{\v n,\v m}.
\end{equation}
By replacing $|\psi\>$ with $\op{G} |\chi\>$ where $\op{G}$ is the Green's function and $\chi$ is the source wavefunction
we obtain
\begin{equation}
  J_{\v n,\v m} = 2 \Im{\op{G} |\chi\>\<\chi| \op{G}\s}_{\v n,\v m}.
\end{equation}
In the standard NEGF formalism it corresponds to $\op{\Sg}^{\text{in}} = |\chi\>\<\chi|$.
When the source is a contact injecting plane waves this formula takes the form \eqref{inscattering-function}-\eqref{phase-factors}.

\bibliographystyle{apsrev}
\bibliography{graphene}

\end{document}